\documentclass[letterpaper,10pt]{scrartcl}

\usepackage{lineno,hyperref}
\modulolinenumbers[5]
\usepackage[utf8]{inputenc}
\usepackage[english]{babel}
\usepackage{amsmath,amssymb,upref}
\usepackage{amsfonts}
\usepackage{amsthm,array,makecell}
\usepackage{color}
\usepackage{graphicx}
\usepackage{sidecap}
\usepackage{multirow}
\usepackage{booktabs}
\usepackage{fullpage}
\usepackage[title]{appendix}
\usepackage{epstopdf}

\usepackage[font=small]{caption}
\usepackage{subcaption}

\usepackage{longtable}
\usepackage{rotating}

\usepackage{algorithm}
\usepackage{algcompatible}

\usepackage{grffile}
\graphicspath{{figures/}}

\usepackage{pgfplots}
\pgfplotsset{width=10cm,compat=1.9}
\usepackage{xcolor}
\definecolor{dark-violet}{RGB}{148,0,211}
\definecolor{sea-green}{RGB}{46, 139,  87}

\DeclareGraphicsExtensions{.pdf,.eps,.png,.jpg,.jpeg}
\newcommand{\R}{\mathbb{R}}

\newcommand{\vb}[1]{\mathbf{#1}}
\newcommand{\lp}{\left(}
\newcommand{\rp}{\right)}
\newcommand{\lb}{\left[}
\newcommand{\rb}{\right]}
\newcommand{\E}{\mathbb{E}}
\newcommand{\Var}{\operatorname{Var}}

\newcommand{\wt}{\widetilde}
\newcommand{\wh}{\widehat}
\def\scriptn{{\mathcal N}}

\newtheorem{theorem}{Theorem}[section]
\newtheorem{proposition}[theorem]{Proposition}
\newtheorem{remark}[theorem]{Remark}

\newcount\Comments  
\newcommand{\kibitz}[2]{\ifnum\Comments=1\textcolor{#1}{#2}\fi}

\ifpdf
\hypersetup{
  pdftitle={Operator inference with roll outs for learning reduced models from scarce and low-quality data}, 
  pdfauthor={Frederick Law and Antoine Cerfon and Benjamin Peherstorfer and Florian Wechsung} 
}
\fi

\newenvironment{keyword}%
   {\begin{trivlist}\item[]{\bfseries\sffamily Keywords:}\ }
   {\end{trivlist}}

   \title{Meta variance reduction for Monte Carlo estimation of energetic particle confinement during stellarator optimization}

\author{Frederick Law\footnote{Courant Institute of Mathematical Sciences, New York University, 251 Mercer Street, New York, NY 10012, USA} \and Antoine Cerfon\footnotemark[1] \and Benjamin Peherstorfer\footnotemark[1] \and Florian Wechsung\footnotemark[1]}

\begin{document}

\maketitle

\begin{abstract}
This work introduces meta estimators that combine multiple multifidelity techniques based on control variates, importance sampling, and information reuse to yield a quasi-multiplicative amount of variance reduction. The proposed meta estimators are particularly efficient within outer-loop applications when the input distribution of the uncertainties changes during the outer loop, which is often the case in reliability-based design and shape optimization. We derive asymptotic bounds of the variance reduction of the meta estimators in the limit of convergence of the outer-loop results. We demonstrate the meta estimators, using data-driven surrogate models and biasing densities, on a design problem under uncertainty motivated by magnetic confinement fusion, namely the optimization of stellarator coil designs to maximize the estimated confinement of energetic particles. The meta estimators outperform all of their constituent variance reduction techniques alone, ultimately yielding two orders of magnitude speedup compared to standard Monte Carlo estimation at the same computational budget.
\end{abstract}


\section{Introduction} \label{section:intro}

\begin{figure}[!ht]
    \centering
    \includegraphics[width=0.6\textwidth]{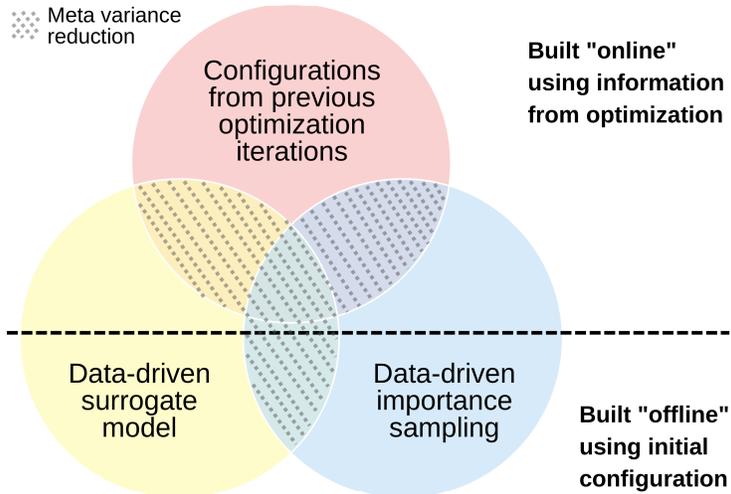}
    \caption{Venn diagram demonstrating the three different estimators to be combined.}
    \label{fig:venn}
\end{figure}

In this article we introduce meta estimators which simultaneously leverage multiple techniques of multifidelity variance reduction to accelerate Monte Carlo estimation. Our meta estimators are based on constituent estimators that each yield variance reduction by taking advantage of different aspects of the estimation problem. First, we build on variance reduction through correlated model outputs via multifidelity Monte Carlo (MFMC) methods with data-driven surrogate models \cite{NgWillcox2014,Peherstorfer2016-MFMC,Peherstorfer2018-MFreview,Gruber2022,KHODABAKHSHI2021}; see also multi-level Monte Carlo methods \cite{doi:10.1287/opre.1070.0496,Cliffe2011,Haji-Ali2016}. 
Second, we combine variance reduction based on control variates with importance sampling (IS) \cite{OwenZhou2000,Zonta2021} with biasing densities that are fitted to data such that they place mass in regions of the input space which we are interested in, akin to multifidelity importance sampling and related techniques \cite{PEHERSTORFER2016490,doi:10.1137/17M1122992,Chaudhuri2020,doi:10.1137/17M1160069,doi:10.1137/19M1257433}. Third, we include the concept of information reuse (IR)  \cite{NgWillcox2014,doi:10.2514/1.C033352,Cook2018} that uses estimators of past optimization iterates as control variates at the current iteration. We contribute a reformulation of IR estimators that guarantees unbiasedness even if distributions of the uncertain inputs change during outer-loop iterations, which is often the case in problems of design under uncertainty. We then show that the proposed meta estimators that combine these three constituent variance reduction techniques are unbiased and asymptotically achieve a quasi-multiplicative amount of variance reduction compared to the constituent estimators. A high-level illustration of this combination process is given in Figure \ref{fig:venn}.

Our application of interest is the uncertainty quantification within the outer loop of stellarator optimization. Stellarators are a promising type of magnetic confinement fusion reactors, which address several of the challenges facing tokamaks, the other strong contenders for commercially viable magnetic confinement fusion energy \cite{sagara2010,helander2012stellarator,Boozer_2021}. Stellarators are complex machines from an engineering point of view, which are designed via lengthy reactor optimization studies based on computationally expensive multi-physics codes \cite{hirshman1998,spong1998,Drevlak_Rose,Lazerson_stellopt,Landreman_simsopt}. One of the key considerations for these design studies is the confinement of energetic alpha particles, born from the fusion of the deuterium and tritium nuclei in the reactor. Good confinement is an essential feature to maximize the net self-heating power \cite{Wolf_2019,Alonso_2022}, and therefore obtain favorable power balance \cite{Alonso_2022,paul2022energetic}. It is also critical to minimize the impact of energetic particle losses on the plasma facing component \cite{sagara2010,ku2008physics,Wolf_2019,Velasco_2021,paul2022energetic}. 

The birth of alpha particles is most accurately described as a random process, where the location of birth of a given particle, and the direction of its initial velocity are random variables. Consequently, alpha particle confinement studies are often done following a Monte Carlo approach \cite{Lotz_1992,Subbotin_2006,henneberg2019properties,albert_2020,LandremanPaulPRL,wechsung2022precise,giuliani_directcomputation,Wechsung_2022}: a large ensemble of initial conditions for alpha particles is generated by sampling the physically appropriate distributions for birth location and velocity direction, and alpha particles trajectories are computed from these initial conditions; one then estimates the alpha particle confinement statistics of interest via the corresponding Monte Carlo estimates obtained from the trajectories. 

Since particle trajectories are expensive to compute for the desired level of accuracy, a direct Monte Carlo approach typically is more computationally expensive than most other physics simulations in stellarator optimization codes, and so costly that it is rarely included in preliminary optimization studies. 
Deterministic measures of the quality of alpha particle confinement have been proposed in the recent past to address this limitation \cite{Nemov2008,bader_2019,Bader_2021,Velasco_2021}, which are less computationally expensive to estimate, and are therefore more practical to include in multi-physics optimization codes. Nevertheless, their predictive capability is imperfect \cite{Bader_2021,Velasco_2021}. Similarly, while a class of stellarator magnetic fields with excellent energetic particle confinement has recently been discovered without targetting this property directly in the optimization process \cite{LandremanPaulPRL,wechsung2022precise,giuliani_directcomputation,landreman2022optimization,Wechsung_2022}, it has not be proven that this remarkable confinement quality would be preserved in more realistic reactor designs, and that this is a robust approach to obtaining good confinement, as we discuss in more detail in Section \ref{section:physics}. The most robust way to achieve good alpha particle confinement remains to include this property as a target in the optimization process, and the most reliable and accurate estimate of alpha particle confinement remains Monte Carlo estimation. It is therefore critical to develop more efficient Monte Carlo estimators, with low enough computational cost enabling their inclusion in multi-physics stellarator optimization codes. This is precisely the goal of the present work.

Variance reduction based on control variates has been used extensively in, e.g., kinetic models such as Boltzmann's equation in \cite{DP19,DP20} and models for micro-turbulence and energetic particle confinement in fusion reactors \cite{Konrad2022,Law2022}. Stochastic collocation techniques based on sparse grids and dimension-adaptive surrogate models for benchmark scenarios of plasma micro-turbulence simulations are introduced in \cite{Fa20,Farca__2021,https://doi.org/10.48550/arxiv.2211.10835}. Other estimation techniques based on polynomial chaos and quasi-Monte Carlo methods are investigated in the context of plasma fusion simulations in \cite{La20,VH18,Di00}. Instead of relying on control variates alone, the proposed meta estimators combine variance reduction with importance sampling, information reuse, and control variates to achieve quasi-multiplicative speedups compared to each of the constituent estimators alone.

This manuscript is structured as follows. In Section \ref{section:prelim} we review the three existing methods we seek to combine, namely: multifidelity Monte Carlo, importance sampling, and information reuse. One of our contributions is in Section \ref{section:AIR} where we adapt the original information reuse and multifidelity information reuse estimators to remain unbiased under changing input distributions, as well as derive asymptotic bounds for their variances. Our primary contribution is in Section \ref{section:combine}, where we combine three multifidelity estimators to construct meta estimators. For meta estimators which have our adaptive information reuse estimator as a constituent method, we also derive the related asymptotic variance reduction. In Section \ref{section:physics} we detail the uncertainty propagation problem of energetic alpha particle confinement in stellarators. In Section \ref{section:numerics} we present our numerical results using meta estimators on a single NCSX-like configuration \cite{Giuliani_nearaxis} and on the optimization trajectory for a new quasi-axisymmetric configuration by Landreman and Paul \cite{LandremanPaulPRL}.

\section{Preliminaries} \label{section:prelim}

We review the standard Monte Carlo, multifidelity Monte Carlo, importance sampling, and information reuse methods. 

\subsection{Standard Monte Carlo estimation} \label{subsection:MC}

We are interested in outer-loop applications under input uncertainty, such as design optimization under uncertainty and robust control. We denote the deterministic outer-loop variable as $\lambda^{(k)} \in \R^{N_{\text{loop}}}, N_{\text{loop}} \in \mathbb{N}$, which depends on the outer-loop iteration $k =0,1,2\dotsc$ and represents, e.g., the design variable and the control parameters. Because we consider outer-loop applications under uncertainty, we also have a domain of input uncertainty that we denote as $D \subset \R^{d}$, which is a simply connected set. 
For each $\lambda^{(k)}$, let $\pi_{\lambda^{(k)}}$ be the probability density function of a probability distribution on $D$, and let $F_{\lambda^{(k)}}:D \to \R$ be a function that represents a high-fidelity model (HFM) of the system of interest.

Although $\pi_{\lambda^{(k)}}$ and $F_{\lambda^{(k)}}$ inherently depend on $\lambda^{(k)}$, since our focus is estimating statistics of $F_{\lambda^{(k)}}$ under $\pi_{\lambda^{(k)}}$ for given $\lambda^{(k)}$, we will suppress this dependence on $\lambda^{(k)}$ in our notation and simply write $\pi^{(k)} = \pi_{\lambda^{(k)}}$ and $F^{(k)} = F_{\lambda^{(k)}}$. Our goal is estimating $\E_{\pi^{(k)}} [F^{(k)}(\xi)]$ at the outer-loop iterations $k = 0, 1, 2, \dots$, where $\xi \sim \pi^{(k)}$ is a random variable on $D$. The computational cost of one evaluation of $F^{(k)}$ is constant for all $k$ and the normalized $\pi^{(k)}$ sampled and evaluated for each $k$. Assume that $\lambda_{k} \to \lambda$ as $k \to \infty$ with $\lambda \in \R^{N_{\text{loop}}}$. Moreover assume $F_{\lambda^{(k)}} \in L^{2}(D)$ for all $k$, and the densities $\pi_{\lambda^{(k)}} \in L^{2}(D)$, with $F_{\lambda^{(k)}} \to F_{\lambda}$ in $L^{2}(D)$ and $\pi_{\lambda^{(k)}} \to \pi_{\lambda}$ in $L^{2}(D)$ which will ensure convergence of the necessary statistics such as expectations, variance, and correlations.

To estimate $\E_{\pi^{(k)}} [F^{(k)}(\xi)]$ with standard Monte Carlo, consider a computational budget of $p$ HFM evaluations at the $k$th outer-loop iteration with samples $\xi_{1}, \dotsc, \xi_{p}$ independent and identically distributed (i.i.d.) from $\pi^{(k)}$. The regular Monte Carlo (MC) estimator and its variance are 
\begin{align}
        \wh{F}^{(k)}_{\text{MC},p} = \frac{1}{p} \sum_{i=1}^{p} F^{(k)}(\xi_{i}), \qquad  \Var_{\pi^{(k)}} \lb \wh{F}^{(k)}_{\text{MC},p} \rb = \frac{\Var_{\pi^{(k)}} [F^{(k)}]}{p}\,, \label{eqn:MC-est}
\end{align}
respectively. 
The MC estimator is unbiased and so the mean-squared error (MSE) is the variance of the estimator. If the variance of $F^{(k)}$ is large, then the number of HFM evaluations $p$ will need to be large in order to make the MSE small and thus attain high accuracy estimation. But if HFM evaluations are expensive, utilizing large $p$ may be too computationally prohibitive. This is true for each outer-loop iteration $k$, and therefore using just the MC estimator alone may be intractable for estimating $\E_{\pi^{(k)}}[F^{(k)}(\xi)]$ during outer-loop applications.

\subsection{Multifidelity Monte Carlo} \label{subsection:MF}

In addition to the HFM $F^{(k)}$, we now also have given a  surrogate model $G^{(k)}: D \to \mathbb{R}$, where we assume $G^{(k)} \in L^{2}(D)$ for all $k$. The multifidelity Monte Carlo (MF) estimator \cite{NgWillcox2014,Peherstorfer2016-MFMC,Peherstorfer18SciTechAIAA,P19AMFMC} leverages $G^{(k)}(\xi)$ as a control variate for estimating the expected value of the HFM $F^{(k)}$. Let $\rho_{\pi^{(k)}}(F^{(k)},G^{(k)})$ be the Pearson's correlation coefficient between $F^{(k)}(\xi)$ and $G^{(k)}(\xi)$ under $\pi^{(k)}$, and let $w(F^{(k)},G^{(k)})$ denote the ratio of cost of evaluating $F^{(k)}$ to the cost for evaluating $G^{(k)}$.  
Then the MF estimator with computational budget equivalent to $p$ HFM evaluations  is
\begin{align}
   \wh{F}^{(k)}_{\text{MF},p} := 
   \operatorname{MF}(F^{(k)},G^{(k)}, \pi^{(k)},p) = \lp \frac{1}{n} \sum_{i=1}^{n} F^{(k)}(\xi_{i}) \rp + \alpha \lp \frac{1}{m} \sum_{i=1}^{m} G^{(k)}(\xi_{i}) - \frac{1}{n} \sum_{i=1}^{n} G^{(k)}(\xi_{i}) \rp \label{eqn:MF-est}
\end{align}
where $\xi_{i}$ are drawn i.i.d. from $\pi^{(k)}$, $n$ is the number of HFM evaluations, $m$ is the number of surrogate model evaluations, and 
\begin{align*}
    \alpha = \rho_{\pi^{(k)}}(F^{(k)},G^{(k)}) \sqrt{\frac{\Var_{\pi^{(k)}}[F^{(k)}]}{\Var_{\pi^{(k)}}[G^{(k)}]}}, \quad p = n + \frac{m}{w}, \quad m = n \sqrt{\frac{w \rho_{\pi^{(k)}}(F^{(k)},G^{(k)})^{2}}{1-\rho_{\pi^{(k)}}(F^{(k)},G^{(k)})^{2}}}\,.
\end{align*}
The variance of the MF estimator is 
\begin{align}
        \Var_{\pi^{(k)}} \lb  \wh{F}^{(k)}_{\text{MF},p} \rb = c_{\text{MF}}(F^{(k)}, G^{(k)}, \pi^{(k)}) \Var_{\pi^{(k)}} \lb \wh{F}^{(k)}_{\text{MC},p} \rb\,, \label{eqn:MF-var}
\end{align}
where $c_{\text{MF}}(F,G,\pi)$ is the variance reduction achieved by the multifidelity estimator using $\operatorname{MF}(F,G,\pi,p)$ compared to the MC estimator for $F$ with $p$ samples, given by
\begin{align*}
    c_{\text{MF}}(F,G,\pi) = \lp \sqrt{1 - \rho_{\pi}(F,G)^{2}} + \sqrt{\frac{\rho_{\pi}(F,G)^{2}}{w(F,G)}} \rp^{2}\,.
\end{align*}
A high variance reduction is achieved if the surrogate model $G^{(k)}$ is cheap to evaluate and its output random variable $G^{(k)}(\xi)$ is highly correlated to the HFM output $F^{(k)}(\xi)$. The MF estimator shrinks the variance by taking advantage of correlation between model \textit{outputs}, and the variance reduction $c_{\text{MF}}(F^{(k)}, G^{(k)}, \pi^{(k)})$ is independent of the computational budget $p$. We note that in the rest of this work we will consider multiple other multifidelity estimators $\operatorname{MF}(F,G,\pi,p)$ using different choices of $F$, $G$, $\pi$ and $p$, with $\alpha$, $n$, $m$, and $w$ implicitly changing correspondingly. 

\subsection{Importance sampling} \label{subsection:IS}

We now review importance sampling that judiciously puts more weight in regions of the model \emph{input} domain with a biasing distribution to reduce the variance. Note that this is different from the MF estimator that achieves variance reduction by taking advantage of correlated model \emph{outputs}. Let $\tilde{\pi}^{(k)}$ be the density of a biasing distribution on $D$ that satisfies $\operatorname{supp}(\pi^{(k)}) \subseteq \operatorname{supp}(\tilde{\pi}^{(k)})$.
Let further $\wt{F}^{(k)}:D \to \R, \xi \mapsto F^{(k)}(\xi) \pi^{(k)}(\xi) / \tilde{\pi}^{(k)}(\xi)$ denote the importance weighted HFM. In the following, the cost of evaluating $F^{(k)}$ and $\wt{F}^{(k)}$ is the same for a given $\xi$, because the cost of evaluating $\pi^{(k)}$ and $\tilde{\pi}^{(k)}$ is typically negligible compared to $F^{(k)}$. 

The IS estimator with a computational budget of $p$ HFM evaluations samples $\xi_{1}, \dotsc, \xi_{p}$ i.i.d. from $\tilde{\pi}^{(k)}$, evaluates $\wt{F}^{(k)}$ for each sample, and takes the sample average:
\begin{align}
    \wh{F}^{(k)}_{\text{IS},p} = \frac{1}{p} \sum_{i=1}^{p} \wt{F}^{(k)} (\xi_{i})\,.
\end{align}
The variance of the IS estimator is
\begin{align}
    \Var_{\tilde{\pi}^{(k)}} \lb \wh{F}^{(k)}_{\text{IS},p} \rb = c_{\text{IS}}^{(k)} \Var_{\pi^{(k)}} \lb \wh{F}^{(k)}_{\text{MC},p} \rb, \qquad c_{\text{IS}}^{(k)} := \frac{\Var_{\tilde{\pi}^{(k)}}[\wt{F}^{(k)}]}{ \Var_{\pi^{(k)}}[F^{(k)}]}\,.
\end{align}
The IS estimator is unbiased, as $\E_{\tilde{\pi}^{(k)}} [\wt{F}^{(k)}] = \E_{\pi^{(k)}}[F^{(k)}]$ provided $(F^{(k)}\pi^{(k)})^{2}/\tilde{\pi}^{(k)} \in L^{1}(D)$. The optimal biasing density is $\tilde{\pi}^{(k)}_{\ast} \propto |F^{(k)}|\pi^{(k)}$ in that $\Var_{\tilde{\pi}_{\ast}^{(k)}}[F \pi^{(k)}/\tilde{\pi}^{(k)}_{\ast}] = 0$. So ideally we would want our biasing $\tilde{\pi}^{(k)}$ to be as close to $\tilde{\pi}^{(k)}_{\ast}$ as possible. Similarly to MF, once the biasing density $\tilde{\pi}^{(k)}$ is chosen, the amount of variance reduction $c_{\text{IS}}^{(k)}$ is fixed for any computational budget $p$.

\subsection{Information reuse estimators} \label{subsection:IR}

Information reuse is an uncertainty quantification concept developed to take advantage of the outer-loop context \cite{NgWillcox2014,doi:10.2514/1.C033352} by leveraging HFM output random variables from previous outer-loop iterations as a control variate at the current iteration. For each outer-loop iteration $k$, the IR estimator introduced in \cite{NgWillcox2014,doi:10.2514/1.C033352} uses the HFM output random variable $F^{(k-1)}(\xi)$ corresponding to outer-loop variable $\lambda^{(k-1)}$ to define a control variate at iteration $k$: with a budget of $p$ HFM evaluations and samples $\xi_{1}, \dotsc, \xi_{p}$ drawn i.i.d. from $\pi^{(k)}$, the IR estimator is
\begin{align*}
        \wh{F}^{(k)}_{\text{IR},p} = \lp \frac{1}{p/2} \sum_{i=1}^{p/2} F^{(k)} (\xi_{i}) \rp + \gamma \lp \wh{F}^{(k-1)}_{\text{IR},p} - \frac{1}{p/2} \sum_{i=1}^{p/2} F^{(k-1)}(\xi_{i}) \rp, \qquad \wh{F}^{(0)}_{\text{IR},p} = \wh{F}^{(0)}_{\text{MC},p}\,,
\end{align*}
where $\gamma$ is a constant which is chosen to minimize the MSE of the estimator. The IR estimator at iteration $k$ reuses both the IR estimator from iteration $k-1$ as well as the HFM $F^{(k-1)}$ from iteration $k - 1$.

Information reuse can also be combined with the multifidelity approach \cite{NgWillcox2014,doi:10.2514/1.C033352,Cook2018} by replacing the MC estimators present in the IR estimator with MF estimators instead. The multifidelity information reuse (MFIR) estimator, using surrogates $G^{(k)}$ and $G^{(k-1)}$ is then

\begin{align*}
    \wh{F}_{\text{MFIR},p}^{(k)} = \operatorname{MF} \lp F^{(k)},G^{(k)},\pi^{(k)},\frac{p}{2} \rp + \gamma \lp \wh{F}_{\text{MFIR},p}^{(k-1)} - \operatorname{MF} \lp F^{(k-1)},G^{(k-1)},\pi^{(k)},\frac{p}{2} \rp \rp
\end{align*}
where the MFIR estimator is initialized with an MF estimator, $\wh{F}_{\text{MFIR},p}^{(0)} = \wh{F}_{\text{MF},p}^{(0)}$ and samples drawn from $\pi^{(k)}$ are the same across both MF estimators. The MFIR estimator at the $k$th outer-loop iteration depends on the MFIR estimator and HFM from iteration $k - 1$ and the surrogates from the $k$th and $(k-1)$st iterations. Similarly to IR, the constant $\gamma$ can be chosen to minimize the MSE of the estimator and depends on the variance of the MFIR estimator at iteration $k - 1$.

\section{Adaptive information reuse estimators}
\label{section:AIR}

In this section we introduce an extension of the IR and MFIR estimators \cite{NgWillcox2014,doi:10.2514/1.C033352,Cook2018} to the case in which the input distribution depends on the outer-loop variable $\lambda^{(k)}$ and thus evolves between iterations. 

\subsection{Adaptive information reuse} \label{subsection:AIR}

In the case of an input distribution that changes with the outer-loop iteration, the IR and MFIR estimators are biased with respect to $\E_{\pi^{(k)}}[F^{(k)}]$. This can be seen from the very first IR iteration:
\begin{align*}
    \E_{\pi^{(1)}}[\wh{F}^{(1)}_{\text{IR},p}] = \E_{\pi^{(1)}}[F^{(1)}] + \gamma (\E_{\pi^{(0)}}[F^{(0)}] - \E_{\pi^{(1)}}[F^{(0)}] )\,,
\end{align*}
which leads to a biased estimator since generically $\E_{\pi^{(0)}}[F^{(0)}] \neq \E_{\pi^{(1)}}[F^{(0)}]$ unless $\lambda^{(0)} = \lambda^{(1)}$, and likewise for the MFIR estimator. The novelty of the adaptive estimators that we now propose is that they correct the bias incurred by the changing input distribution.

\subsubsection{Adaptive information reuse (AIR) estimator}

We introduce the adaptive information reuse (AIR) estimator with budget $p$ as \begin{align}
    \wh{F}^{(k)}_{\text{AIR},p} = \lp \frac{1}{p/2} \sum_{i=1}^{p/2} F^{(k)} (\xi_{i}) \rp + \gamma \lp \wh{F}^{(k-1)}_{\text{AIR},p} - \frac{1}{p/2} \sum_{i=1}^{p/2} F^{(k-1)}(\xi_{i}) W_{k}(\xi_{i}) \rp\,, \label{eqn:AIR-est}
\end{align}
where $k = 1, 2, \dots$ is the current outer-loop iteration and $\wh{F}^{(0)}_{\text{AIR},p} =  \wh{F}^{(0)}_{\text{MC},p}$. The samples $\xi_{1},\dotsc,\xi_{p}$ are drawn i.i.d.~from $\pi^{(k)}$. The weight $W_k(\xi)$ is
\begin{align}
    W_{k}(\xi) = \frac{\pi^{(k-1)}(\xi)}{\pi^{(k)}(\xi)}\,.
\end{align}

We now prove that this estimator is indeed unbiased for all $k$ as long as $\wh{F}^{(0)}_{\text{AIR},p}$ is unbiased in the sense that $\E_{\pi^{(0)}}[\wh{F}^{(0)}_{\text{AIR},p}] = \E_{\pi^{(0)}}[F^{(0)}]$.
\begin{proposition} \label{prop:unbiasedIR}
Suppose we initialize the AIR estimator with an MC estimator using a budget of $p$ HFM evaluations, $\wh{F}^{(0)}_{\text{AIR},p} = \wh{F}^{(0)}_{\text{MC},p}$. Then, in each iteration of the outer loop $k = 1, 2, 3, \dots$, the iteratively defined estimator $\wh{F}^{(k)}_{\text{AIR},p}$, is unbiased, which means that $\E_{\pi^{(k)}}[\wh{F}^{(k)}_{\text{AIR},p}] = \E_{\pi^{(k)}}[F^{(k)}]$ holds.
\end{proposition}
\begin{proof}
The proof is by induction. Since the MC estimator is unbiased and the AIR estimator is initialized as an MC estimator, we have that $\E_{\pi^{(0)}}[\wh{F}^{(0)}_{\text{AIR},p}] = \E_{\pi^{(0)}}[F^{(0)}]$. For $k > 0$, suppose that $\E_{\pi^{(k-1)}}[\wh{F}^{(k-1)}_{\text{AIR},p}] = \E_{\pi^{(k-1)}}[F^{(k-1)}]$. By the law of total expectation:
\begin{align*}
     \E_{\pi^{(k)}}[\wh{F}^{(k)}_{\text{AIR},p}] &= \E_{\pi^{(k-1)}}[\E_{\pi^{(k)}}[\wh{F}^{(k)}_{\text{AIR},p} | \wh{F}^{(k-1)}_{\text{AIR},p}]] = \E_{\pi^{(k-1)}} \lb \E_{\pi^{(k)}}[F^{(k)}] + \gamma (\wh{F}^{(k-1)}_{\text{AIR},p} - \E_{\pi^{(k)}}[F^{(k-1)}W_{k}] ) \rb\\
    &= \E_{\pi^{(k-1)}} \lb \E_{\pi^{(k)}}[F^{(k)}] + \gamma \lp \wh{F}^{(k-1)}_{\text{AIR},p} - \E_{\pi^{(k)}} \lb F^{(k-1)} \frac{\pi^{(k-1)}}{\pi^{(k)}} \rb \rp \rb\\
    &= \E_{\pi^{(k-1)}} \lb \E_{\pi^{(k)}}[F^{(k)}] - \gamma ( \wh{F}^{(k-1)}_{\text{AIR},p} - \E_{\pi^{(k-1)}} [F^{(k-1)}] ) \rb\\
    &=  \E_{\pi^{(k)}}[F^{(k)}] - \gamma ( \E_{\pi^{(k-1)}}[\wh{F}^{(k-1)}_{\text{AIR},p}] - \E_{\pi^{(k-1)}} [F^{(k-1)}] ) = \E_{\pi^{(k)}}[F^{(k)}]
\end{align*}
where the last step uses the inductive hypothesis that $\E_{\pi^{(k-1)}}[\wh{F}^{(k-1)}_{\text{AIR},p}] = \E_{\pi^{(k-1)}}[F^{(k-1)}]$. Thus we deduce that $\E_{\pi^{(k)}}[\wh{F}^{(k)}_{\text{AIR},p}] = \E_{\pi^{(k)}}[F^{(k)}]$ holds for all $k$, and the AIR estimator is unbiased.
\end{proof}

\subsubsection{Variance of the AIR estimator}
The $\gamma^*_k$ that minimizes the variance of the AIR estimator at iteration $k$ can be derived analogously to the case of the IR estimator with an outer-loop iteration-independent input distribution \cite{NgWillcox2014}. It is 
\begin{align*}
    \gamma^{\ast}_{k,\text{AIR}} = \lp \frac{\rho_{\pi^{(k)}}(F^{(k)},F^{(k-1)}W_{k})}{1 + \eta_{k}} \rp \sqrt{\frac{\Var_{\pi^{(k)}}[F^{(k)}]}{\Var_{\pi^{(k)}}[F^{(k-1)}W_{k}]}}, \qquad  \eta_{k} =  \frac{\Var_{\pi^{(k-1)}}[\wh{F}^{(k-1)}_{\text{AIR},p}]}{\Var_{\pi^{(k)}}[F^{(k-1)}W_{k}] / (p/2)}
\end{align*}
and the variance of the AIR estimator with $\gamma^{\ast}_{k,\text{AIR}}$ is
\begin{align}
    \Var_{\pi^{(k)}}[\wh{F}^{(k)}_{\text{AIR},p}] = c_{\text{AIR}}^{(k)} \Var_{\pi^{(k)}}[\wh{F}_{\text{MC},p}^{(k)}], \qquad c_{\text{AIR}}^{(k)} = 2 \lp 1 - \frac{\rho_{\pi^{(k)}}(F^{(k)},F^{(k-1)}W_{k})^{2}}{1+\eta_{k}} \rp \label{eqn:AIR-var}
\end{align}
where $c_{\text{AIR}}^{(k)}$ is the variance reduction using the AIR estimator compared to the MC estimator with equivalent computational cost, at the $k$th outer-loop iteration. To compute $c_{\text{AIR}}^{(k)}$ we must compute $\eta_{k}$ which depends on $\Var_{\pi^{(k-1)}}[\wh{F}_{\text{AIR},p}^{(k-1)}]$ which itself depends on $c_{\text{AIR}}^{(k-1)}$. But then $c_{\text{AIR}}^{(k-1)}$ depends on $c_{\text{AIR}}^{(k-2)}$ in the same manner, and so on recursively. 

A disadvantage of this recurrence relation is that in order to compute the variance reduction using AIR at the $k$th outer-loop iteration, we must have already run the AIR estimator at all $l$th outer-loop iterations for $0\leq l \leq k$. To avoid this recurrence, we now derive an asymptotic approximation of the variance reduction $c_{\text{AIR}}^{(k)}$ which can be computed at the $k$th outer-loop iteration which does not require information from all previous iterations. We thus consider the limit of $c_{\text{AIR}}^{(k)}$ with respect to $k \to \infty$, which we expect to be a good approximation of  $c_{\text{AIR}}^{(k)}$ for large $k$. The following proposition derives the variance reduction of the AIR estimator for $k \to \infty$. 

\begin{proposition} \label{prop:IR-asym}
Suppose that variance reduction $c_{\text{AIR}}^{(k)}$ using the AIR estimator converges to $c_{\text{AIR}} \geq 0$ as $k \to \infty$. Moreover, suppose that $\rho_{\pi^{(k)}}(F^{(k)},F^{(k-1)}W_{k})^{2} \to \rho^{2}$ as $k \to \infty$ so that the correlation between current high-fidelity and reweighted previous high-fidelity converges. Finally, suppose that the ratio of variances $\Var_{\pi^{(k-1)}}[F^{(k-1)}] / \Var_{\pi^{(k)}} [F^{(k-1)}W_{k}] \to 1$ as $k \to \infty$. Then the asymptotic variance reduction is given by $c_{\text{AIR}} = 2 \sqrt{1-\rho^{2}}$.
\end{proposition}
\begin{proof}
Under the assumption that $\Var_{\pi^{(k-1)}}[F^{(k-1)}] / \Var_{\pi^{(k)}} [F^{(k-1)}W_{k}] \to 1$ as $k \to \infty$, we see that
\begin{align*}
    \eta_{k} = \frac{\Var_{\pi^{(k-1)}}[\wh{F}^{(k-1)}_{\text{AIR},p}]}{\Var_{\pi^{(k)}}[F^{(k-1)}W_{k}] / (p/2)} = \frac{c_{\text{AIR}}^{(k-1)} \Var_{\pi^{(k-1)}}[F^{(k-1)}]/p}{\Var_{\pi^{(k)}}[F^{(k-1)}W_{k}]/(p/2)} \to \frac{c_{\text{AIR}}}{2} \quad \text{as } k \to \infty
\end{align*}
Dividing the equation for $c_{\text{AIR}}^{(k)}$ (\ref{eqn:AIR-var}) by $2$, and passing to the limit we get
\begin{align*}
    \frac{c_{\text{AIR}}^{(k)}}{2} = 1 - \frac{\rho_{\pi^{(k)}}(F^{(k)},F^{(k-1)}W_{k})^{2}}{1 + \eta_{k}} \quad \overset{k \to \infty}{\longrightarrow} 
    \quad  \frac{c_{\text{AIR}}}{2} = 1 - \frac{\rho^{2}}{1 + (c_{\text{AIR}}/2)}
\end{align*}
This system can then be solved,
\begin{align*}
    \lp \frac{c_{\text{AIR}}}{2} - 1 \rp \lp \frac{c_{\text{AIR}}}{2} + 1 \rp = - \rho^{2} \quad &\Longrightarrow \quad \frac{c_{\text{AIR}}^{2}}{4} = 1- \rho^{2} \\
    &\Longrightarrow \quad c_{\text{AIR}} = 2 \sqrt{1 - \rho^{2}}
\end{align*}
\end{proof}
The results of Proposition \ref{prop:IR-asym} motivates approximating the variance reduction of the AIR estimator at an iteration $k$ as
\begin{align}
    \hat{c}_{\text{AIR}}^{(k)} = \hat{c}_{\text{AIR}}(F^{(k)},F^{(k-1)}, \pi^{(k)}, \pi^{(k-1)})
\end{align}
where for $\mathcal{F}_{1}, \mathcal{F}_{2} \in L^{2}(D)$ and $\mu,\nu$ probability densities on $D$, $\hat{c}_{\text{AIR}}(\mathcal{F}_{1}, \mathcal{F}_{2}, \mu, \nu)$ is defined by
\begin{align}
    \hat{c}_{\text{AIR}}(\mathcal{F}_{1}, \mathcal{F}_{2}, \mu, \nu) := 2 \sqrt{1 - \rho_{\mu}\lp \mathcal{F}_{1},\mathcal{F}_{2} \frac{\nu}{\mu} \rp^{2}}. \label{eqn:AIR-var-asym}
\end{align}
By Proposition \ref{prop:IR-asym}, $|\hat{c}_{\text{AIR}}^{(k)} - c_{\text{AIR}}^{(k)}| \to 0$ as $k \to \infty$. A major advantage of $\hat{c}_{\text{AIR}}^{(k)}$ is that unlike equation (\ref{eqn:AIR-var}), it is independent of any estimator used at the previous steps. At each outer-loop iteration $k=1,2,\dotsc$, the approximate variance $\hat{c}_{\text{AIR}}^{(k)}$ depends only on $F^{(k)}$, $F^{(k-1)}$, $\pi^{(k)}$, and $\pi^{(k-1)}$. Thus, the values $F^{(k)}(\xi_{i})$, $F^{(k-1)}(\xi_{i})$, $W_{k}(\xi_{i})$ for $i=1,\dotsc,p/2$ used to construct the AIR estimator (\ref{eqn:AIR-est}) can be reused to compute the sample correlation coefficient which can then be used in equation (\ref{eqn:AIR-var-asym}) to estimate $\hat{c}_{\text{AIR}}^{(k)}$.

\begin{remark}
The assumptions required in deriving $\hat{c}_{\text{AIR}}^{(k)}$ are satisfied assuming the outer-loop variables $\lambda^{(k)}$ to converge to $\lambda$ as $k \to \infty$, and the models and distributions depend on the outer-loop variable continuously. Given our assumptions on the $L^{2}(D)$ convergence of $\pi_{\lambda^{(k)}}$ and $F_{\lambda^{(k)}}$ depend continuously on $\lambda^{(k)}$, this guarantees that the necessary correlations and variance ratios will converge as well.
\end{remark}

\subsection{Multifidelity adaptive information reuse} \label{subsection:MFAIR}

We consider the MFIR estimator and correct the bias with a weight if the input distribution $\pi^{(k)}$ changes with the outer-loop iterations. 

\subsubsection{Multifidelity adaptive information reuse estimator (MFAIR)}
The multifidelity adaptive information reuse (MFAIR) estimator with a computational budget of $p$ is
\begin{align}
    \wh{F}_{\text{MFAIR},p}^{(k)} =  
    \operatorname{MF} \lp F^{(k)}, G^{(k)}, \pi^{(k)}, \frac{p}{2} \rp
     + \gamma \lb \wh{F}_{\text{MFAIR},p}^{(k-1)} -
     \operatorname{MF} \lp F^{(k-1)} W_{k}, G^{(k-1)}, \pi^{(k)}, \frac{p}{2} \rp \rb \label{eqn:MFIR-est}
\end{align}
where we initialize the MFAIR estimator with an MF estimator, $ \wh{F}_{\text{MFAIR},p}^{(0)} = \wh{F}_{\text{MF},p}^{(0)}$. Similarly to the AIR estimator, the samples of $\xi$ drawn i.i.d. from $\pi^{(k)}$ to form $\operatorname{MF}( F^{(k)}, G^{(k)}, \pi^{(k)}, p/2 )$ are the same samples as those used in $\operatorname{MF} ( F^{(k-1)} W_{k}, G^{(k-1)}, \pi^{(k)}, p/2)$. This is necessary for the samples $F^{(k)}(\xi_{i})$, $G^{(k)}(\xi_{i})$, $F^{(k-1)}(\xi_{i})W_{k}(\xi_{i})$, $G^{(k-1)}(\xi_{i})$ to be correlated. Subsequently the MF estimators appearing in the MFAIR estimator (\ref{eqn:MFIR-est}) are correlated.

The MFAIR estimator is unbiased, which can be shown with similar arguments as used in the proof of Proposition \ref{prop:unbiasedIR}. Because of the unbiasedness, the MSE of the MFAIR is its variance. One option to find a good balancing parameter $\gamma$ is to minimize the variance, which leads to the balancing parameter
\begin{align}\label{eq:MFAIRVar}
    \gamma^{\ast}_{k,\text{MFAIR}} = \frac{\operatorname{COV}_{k}}{\Var_{\pi^{(k-1)}}[\wh{F}_{\text{MFAIR},p}^{(k-1)}] + \Var_{\pi^{(k)}}[\operatorname{MF} ( F^{(k-1)} W_{k}, G^{(k-1)}, \pi^{(k)}, p/2)]} 
\end{align}
where the term $\operatorname{COV}_{k}$ is the covariance of $\operatorname{MF}( F^{(k)}, G^{(k)}, \pi^{(k)}, p/2 )$ and $\operatorname{MF} ( F^{(k-1)} W_{k}, G^{(k-1)}, \pi^{(k)}, p/2)$
under the distribution $\pi^{(k)}$,
\begin{equation}
\begin{aligned}
    \operatorname{COV}_{k} = C_{k} \frac{\sqrt{\Var_{\pi^{(k)}}[F^{(k)}] \Var_{\pi^{(k)}}[F^{(k-1)}W_{k}]}}{n} \label{eqn:MFIR-cov}
\end{aligned}
\end{equation}
where the coefficient $C_{k}$ depends on the $\binom{4}{2}$ possible correlations between HFMs and surrogates,
\begin{equation*}
\begin{aligned}
    C_{k} =& \bigg[ \rho_{\pi^{(k)} } (F^{(k)},F^{(k-1)}W_{k}) + \lp 1 - \frac{1}{r^{\ast}_{k}} \rp \bigg( \rho_{\pi^{(k)}}(F^{(k)},G^{(k)}) \rho_{\pi^{(k)}}(F^{(k-1)}W_{k},G^{(k-1)}) \rho_{\pi^{(k)}}(G^{(k)},G^{(k-1)}) \\
    &-\rho_{\pi^{(k)}}(F^{(k)},G^{(k)}) \rho_{\pi^{(k)}}(F^{(k-1)}W_{k},G^{(k)}) - \rho_{\pi^{(k)}}(F^{(k-1)}W_{k},G^{(k-1)}) \rho_{\pi^{(k)}}(F^{(k)},G^{(k-1)}) \bigg) \bigg].
\end{aligned}
\end{equation*}
The $r_{k}^{\ast}$ term is given by
\begin{align*}
    r^{\ast}_{k} = \frac{m}{n} = \min \lp \sqrt{\frac{w(F^{(k)},G^{(k)}) \rho_{\pi^{(k)}}(F^{(k)},G^{(k)})^{2}}{1 - \rho_{\pi^{(k)}}(F^{(k)},G^{(k)})^{2}}}, \sqrt{\frac{w(F^{(k-1)},G^{(k-1)}) \rho_{\pi^{(k)}}(F^{(k-1)}W_{k},G^{(k-1)})^{2}}{1 - \rho_{\pi^{(k)}}(F^{(k-1)}W_{k},G^{(k-1)})^{2}}} \rp
\end{align*}
where $n$ is the number of HFM evaluations $m$ is the number of low-fidelity model evaluations, satisfying $p/2 = n + m/w$. The variance of the MFAIR estimator with balancing parameter \eqref{eq:MFAIRVar} is
\begin{align}
    \Var_{\pi^{(k)}} [\wh{F}_{\text{MFAIR},p}^{(k)}] = \Var_{\pi^{(k)}} [\wh{F}_{\text{MF},p/2}^{(k)}] - \frac{\operatorname{COV}_{k}^{2}}{\Var_{\pi^{(k-1)}}[\wh{F}_{\text{MFAIR},p}^{(k-1)}] + \Var_{\pi^{(k)}}[\wh{F}_{\text{MF},F^{(k-1)}W_{k},G^{(k-1)},\pi^{(k)},p/2}]} \label{eqn:MFAIR-var}
\end{align}

\subsubsection{Asymptotic variance reduction of MFAIR}
We now derive the asymptotic expression of the variance reduction of the MFAIR estimator compared to the MC estimator, under the additional assumption that the surrogate models converge $G^{(k)} \to G$ as $k \to \infty$ in $L^{2}(D)$. Note that if the surrogate model $G$ is fixed $G^{(k)} = G$ for $k=1,2,\dotsc$, then this condition is satisfied as well.

Let $c_{\text{MFAIR}}^{(k)}$ be the variance reduction using $\wh{F}_{\text{MFAIR},p}^{(k)}$ with $\gamma^{\ast}_{k,\text{MFAIR}}$, compared to $\wh{F}_{\text{MFAIR},p}^{(k)}$,
\begin{align*}
    c_{\text{MFAIR}}^{(k)} = \Var_{\pi^{(k)}} [\wh{F}_{\text{MFAIR},p}^{(k)}]/\Var_{\pi^{(k)}}[\wh{F}_{\text{MC},p}^{(k)}]\,.
\end{align*}
Just as the variance reduction $c_{\text{AIR}}^{(k)}$ depends recursively on the HFM and input distribution history at all previous outer-loop iterations, the variance reduction $c_{\text{MFAIR}}^{(k)}$ depends on the $F^{(l)}$, $G^{(l)}$, and $\pi^{(l)}$ for all $l \leq k$. And just as for the AIR estimator, a disadvantage of this recurrence relation is that to compute $c_{\text{MFAIR}}^{(k)}$, one must have already computed $c_{\text{MFAIR}}^{(l)}$ for $l \leq k$, which requires running the MFAIR estimator for all previous outer-loop iterations. To provide flexibility to the practitioner, we similarly derive an asymptotic approximation for the variance reduction $c_{\text{MFAIR}}^{(k)}$ which can be computed at the $k$th outer-loop iteration without requiring information from all previous iterations. We consider the limit of $c_{\text{MFAIR}}^{(k)}$ when $k \to \infty$, and expect this to be an accurate approximation of $c_{\text{MFAIR}}^{(k)}$ for large $k$. The following proposition derives the variance reduction of the MFAIR estimator for $k \to \infty$.

\begin{proposition} \label{prop:MFIR-asym}
Suppose that $c_{\text{MFAIR}}^{(k)} \to c_{\text{MFAIR}} \geq 0$ as $k \to \infty$. Additionally, suppose that the correlations between HFMs and surrogates converge to the same value for the outer-loop iterations $k \to \infty$, so that $\rho_{\pi^{(k)}}(F^{(k)},G^{(k)})$, $\rho_{\pi^{(k)}}(F^{(k)}, G^{(k-1)})$, $\rho_{\pi^{(k)}}(F^{(k-1)}W_{k},G^{(k)})$, and $\rho_{\pi^{(k)}}(F^{(k-1)}W_{k},G^{(k-1)})$ all converge to $\rho$ as $k \to \infty$ for $\rho < 1$. Moreover, suppose the surrogates $G^{(k)}$ satisfy $\rho_{\pi^{(k)}}(G^{(k)},G^{(k-1)}) \to 1$ as $k \to \infty$.
Finally, suppose that the correlation between the HFM output random variable at the current and the previous iteration converge so that $\rho_{\pi^{(k)}}(F^{(k)},F^{(k-1)}W_{k}) \to \hat{\rho}$ as $k \to \infty$ and  that $\Var_{\pi^{(k-1)}}[F^{(k-1)}]/\Var_{\pi^{(k)}}[F^{(k-1)}W_{k}] \to 1$ as $k \to \infty$.
Then
\begin{align*}
    c_{\text{MFAIR}} = 2 \lp 1 + \frac{r^{\ast}}{w(F,G)} \rp \sqrt{\lp 1- \lp 1 - \frac{1}{r^{\ast}} \rp \rho^{2} \rp^{2} - \lp \hat{\rho}-\lp 1 - \frac{1}{r^{\ast}} \rp \rho^{2} \rp^{2}} 
\end{align*}
where $r^{\ast} = \lim_{k \to \infty} r_{k}^{\ast}$.
\end{proposition}
\begin{proof}
Let $V_{k} = c_{\text{MFAIR}}^{(k)} / [2 (1 +( r_{k}^{\ast}/w(F_{k},G_{k})))]$, so that $\Var_{\pi^{(k)}} [\wh{F}_{\text{MFAIR},p}^{(k)}] = V_{k} \Var_{\pi^{(k)}}[F^{(k)}]/n$. We can express the variance of the two MF estimators \cite{NgWillcox2014} within the MFAIR estimator as
\begin{align*}
    \Var_{\pi^{(k)}}[\wh{F}^{(k)}_{\text{MF},p/2}] &= \lb 1 - \lp 1 - \frac{1}{r^{\ast}_{k}} \rp \rho_{\pi^{(k)}}(F^{(k)},G^{(k)})^{2} \rb \frac{\Var_{\pi^{(k)}}[F^{(k)}]}{n}\,,\\
    \Var_{\pi^{(k)}}[\wh{F}_{\text{MF},F^{(k-1)}W_{k},G^{(k-1)},\pi^{(k)},p/2}] &= \lb 1 - \lp 1 - \frac{1}{r^{\ast}_{k}} \rp \rho_{\pi^{(k)}}(F^{(k-1)}W_{k},G^{(k-1)})^{2} \rb \frac{\Var_{\pi^{(k)}}[F^{(k-1)}W_{k}]}{n}.
\end{align*}
Define $\varphi_{k}^{\ast} = 1 - (1/r^{\ast}_{k})$. Then based on our assumptions of convergence, $\varphi_{k}^{\ast} \to \varphi^{\ast}$ for some $\varphi^{\ast} > 0$. 

Plugging the covariance term (\ref{eqn:MFIR-cov}) and MF variance equations into equation (\ref{eqn:MFAIR-var}) and dividing both sides by $\Var_{\pi^{(k)}}[F^{(k)}]/n$, we get
\begin{multline*}
    V_{k} = \lp 1 - \varphi_{k}^{\ast} \rho_{\pi^{(k)}}(F^{(k)},G^{(k)})^{2} \rp\\ - \frac{C_{k}^{2}}{V_{k-1} \lp \Var_{\pi^{(k-1)}}[F^{(k-1)}] / \Var_{\pi^{(k)}}[F^{(k-1)}W_{k}] \rp + (1-\varphi_{k}^{\ast} \rho_{\pi^{(k)}}(F^{(k-1)}W_{k},G^{(k-1)})^{2})}\,.
\end{multline*}
Under our assumptions, $C_{k} \to \hat{\rho} - \varphi^{\ast} \rho^{2}$ as $k \to \infty$. Thus passing to the limit for the $V_{k}$ relation we get
\begin{align*}
    V = (1-\varphi^{\ast}\rho^{2}) - \frac{(\hat{\rho} - \varphi^{\ast} \rho^{2})^{2}}{V + (1-\varphi^{\ast}\rho^{2})}
\end{align*}
Simplifying and solving for $V$ we get
\begin{align*}
    (V - (1-\varphi^{\ast}\rho^{2}) ) (V + (1-\varphi^{\ast}\rho^{2}) )  = -(\hat{\rho} - \varphi^{\ast}\rho^{2})  \quad \Longrightarrow \quad V = \sqrt{(1-\varphi^{\ast}\rho^{2})^{2} - (\hat{\rho} - \varphi^{\ast}\rho^{2})^{2}}\,.
\end{align*}
Under our assumptions, we will also have $r_{k}^{\ast} \to r^{\ast}$ and thus
\begin{align*}
     c_{\text{MFAIR}} = 2 \lp 1 + \frac{r^{\ast}}{w} \rp \sqrt{(1-\varphi^{\ast}\rho^{2})^{2} - (\hat{\rho} - \varphi^{\ast}\rho^{2})^{2}}\,.
\end{align*}

\end{proof}
Analogous to the case for the AIR estimator, Proposition \ref{prop:MFIR-asym} motivates approximating the variance reduction using the MFAIR estimator at an iteration $k$ by
\begin{equation}
\begin{aligned}
    \hat{c}_{\text{MFAIR}}^{(k)} &= c_{\text{MFAIR}}(F^{(k)}, F^{(k-1)}, G^{(k)}, \pi^{(k)}, \pi^{(k-1)})
\end{aligned} \label{eqn:MFAIR-var-asym}
\end{equation}
where for $\mathcal{F}_{1}, \mathcal{F}_{2}, \mathcal{G} \in L^{2}(D)$ and $\mu,\nu$ probability densities on $D$, $\hat{c}_{\text{MFAIR}}(\mathcal{F}_{1}, \mathcal{F}_{2}, \mathcal{G}, \mu, \nu)$ is defined by
\begin{align*}
    \hat{c}_{\text{MFAIR}}(\mathcal{F}_{1}, \mathcal{F}_{2}, \mathcal{G}, \mu, \nu) :=& \, 2 \lp 1 + \frac{r^{\ast}(\mathcal{F}_{1}, \mathcal{G})}{w(\mathcal{F}_{1}, G)} \rp  \times\\
    &\sqrt{\lp 1 - \varphi^{\ast}(\mathcal{F}_{1}, \mathcal{G}) \rho_{\mu}(\mathcal{F}_{1}, \mathcal{G}) \rp^{2} - \lp \rho_{\mu} \lp \mathcal{F}_{1}, \mathcal{F}_{2} \frac{\nu}{\mu} \rp - \varphi^{\ast}(\mathcal{F}_{1}, \mathcal{G})  \rho_{\mu} (\mathcal{F}_{1}, \mathcal{G})^{2} \rp^{2} },\\
    r^{\ast}(\mathcal{F}_{1}, \mathcal{G}) :=& \,  \sqrt{\frac{w(\mathcal{F}_{1},\mathcal{G}) \rho_{\mu}(\mathcal{F}_{1}, \mathcal{G})^{2}}{1 - \rho_{\mu}(\mathcal{F}, \mathcal{G})^{2}}}, \qquad \varphi^{\ast}(\mathcal{F}_{1}, \mathcal{G}) := 1 - \frac{1}{r^{\ast}(\mathcal{F}_{1}, \mathcal{G})}.
\end{align*}
By Proposition \ref{prop:MFIR-asym}, $|\hat{c}_{\text{MFAIR}}^{(k)} - c_{\text{MFAIR}}^{(k)} | \to 0$ as $k \to \infty$.
As for $\hat{c}_{\text{AIR}}^{(k)}$ for the AIR estimator, $\hat{c}_{\text{MFAIR}}^{(k)}$ can be estimated independently of estimators of previous iterations.

The variance reduction using the MFAIR estimator increases when the correlation between HFMs $\rho_{\pi^{(k)}}(F^{(k)},F^{(k-1)W_{k}})$ gets closer to 1. Moreover, we see that this variance reduction will further increase if the correlation between the high-fidelity and surrogate models $\rho_{\pi^{(k)}}(F^{(k)},G)$ gets closer to 1. Since 
$\hat{c}_{\text{MFAIR}}^{(k)}$ can be measured during the outer loop, this allows us to monitor the efficacy of the MFAIR estimator compared to other estimators.

\section{Boosting variance reduction with meta estimators} \label{section:combine}

In this section we introduce three meta estimators that combine the MF, IS, and AIR estimators from the previous sections. We show that the meta estimators' variance reduction scales in a multiplicative way with the variance reduction of each individual estimator.

\subsection{The ISMF meta estimator: Combining importance sampling and multifidelity Monte Carlo estimation}

Combining importance sampling and multifidelity Monte Carlo estimation leads to the ISMF estimator
\begin{align}
    \wh{F}^{(k)}_{\text{ISMF},p} = \wh{F}_{\text{MF},\wt{F}^{(k)},\wt{G}^{(k)},\tilde{\pi}^{(k)},p} \label{eqn:ISMF-est}
\end{align}
with budget $p$, where the samples $\xi_{i}$ are drawn from the biasing distribution with density $\tilde{\pi}^{(k)}$ instead of the nominal density $\pi^{(k)}$. Similarly to the reweighted HFM output random variable $\wt{F}^{(k)}$, we use the reweighted surrogate output random variable $\wt{G}^{(k)} = G^{(k)} \pi^{(k)}/\tilde{\pi}^{(k)}$ with the importance weights $\pi^{(k)}/\tilde{\pi}^{(k)}$. Thus, the proposed ISMF meta estimator combines the IS and the MF estimator by replacing the MC estimators in the MF estimator \eqref{eqn:MF-est} with their respective IS estimators.  

The variance of the ISMF estimator is
\begin{align}
        \Var_{\tilde{\pi}^{(k)}} \lb  \wh{F}^{(k)}_{\text{ISMF},p} \rb = c_{\text{ISMF}}^{(k)} \Var_{\pi^{(k)}} [\wh{F}_{\text{MC},p}^{(k)}], \quad c_{\text{ISMF}}^{(k)} = c_{\text{MF}}(\wt{F}^{(k)}, \wt{G}^{(k)}, \tilde{\pi}^{(k)}) \times c_{\text{IS}}^{(k)} \label{eqn:ISMF-var}
\end{align}
using $\Var_{\tilde{\pi}^{(k)}}[\wt{F}^{(k)}] = c_{\text{IS}}^{(k)} \Var_{\pi^{(k)}}[F^{(k)}]$.
The variance reduction of the MFIS  estimator depends on the correlation between $\wt{F}^{(k)}$ and $\wt{G}^{(k)}$ under $\tilde{\pi}^{(k)}$ and on $\tilde{w}$, the ratio of the cost of evaluating the HFM to the cost of the evaluating the surrogate when the input samples are drawn from $\tilde{\pi}^{(k)}$. Thus the ISMF estimator provides a quasi-multiplicative variance reduction, in that $c_{\text{ISMF}^{(k)}}$ is the product of variance reduction from an MF estimator using $\wt{F}^{(k)}$, $\wt{G}^{(k)}$, $\tilde{\pi}^{(k)}$ instead of $F^{(k)}$, $G^{(k)}$, $\pi^{(k)}$, with the variance reduction of the IS estimator. So long as $c_{\text{MF}}(\wt{F}^{(k)}, \wt{G}^{(k)}, \tilde{\pi}^{(k)})$ is not too large compared to $c_{\text{MF}}(F^{(k)},G^{(k)},\pi^{(k)})$, we can possibly expect ISMF to outperform both the MF and IS estimators on their own.

\subsection{The ISAIR meta estimator: Combining importance sampling and adaptive information reuse}

For a budget of $p$ HFM evaluations, we define the ISAIR estimator as
\begin{align}
    \wh{F}^{(k)}_{\text{ISAIR},p} = \lp \frac{1}{p/2} \sum_{i=1}^{p/2} \wt{F}^{(k)} (\xi_{i}) \rp + \gamma \lp \wh{F}^{(k-1)}_{\text{ISAIR},p} - \frac{1}{p/2} \sum_{i=1}^{p/2} \wt{F}^{(k-1)}(\xi_{i}) W_{k}(\xi_{i}) \rp\,, \label{eqn:ISIR-est}
\end{align}
which combines importance sampling and adaptive information reuse. The samples $\xi_{1}, \dots, \xi_{p/2}$ are drawn i.i.d.~from $\tilde{\pi}^{(k)}$. The optimal choice of $\gamma$ is functionally the same as $\gamma_{k,\text{AIR}}^{\ast}$, but with $\tilde{\pi}^{(k)}$ replacing $\pi^{(k)}$ and $\wt{F}^{(k)}$ replacing $F^{(k)}$. The variance reduction of the ISAIR estimator $\wh{F}_{\text{ISAIR},p}^{(k)}$ using $\gamma^{\ast}_{k,\text{AIR}}$, compared to $\wh{F}_{\text{MC},p}^{(k)}$ is
\begin{align}
    c_{\text{ISAIR}}^{(k)} = \Var_{\pi^{(k)}}[\wh{F}_{\text{ISAIR},p}^{(k)}] / \Var_{\pi^{(k)}}[\wh{F}_{\text{MC},p}^{(k)}],
\end{align}
which has the same recursive dependence as $c_{\text{AIR}}^{(k)}$, in that $c_{\text{ISAIR}}^{(k)}$ depends on $c_{\text{ISAIR}}^{(k-1)}$ which itself depends on $c_{\text{ISAIR}}^{(k-2)}$ and so on. However, similar to the AIR estimator, we can derive an approximation for the asymptotic ISAIR variance reduction analogous to (\ref{eqn:AIR-var-asym}). The derivation of this approximation follows from the proof of Proposition \ref{prop:IR-asym} but with the necessary models and distributions substituted, assuming that the variance reduction from IS alone $c_{\text{IS}}^{(k)}$ also converges as $k \to \infty$.
Then the asymptotic variance reduction approximation is
\begin{align}
   \hat{c}_{\text{ISAIR}}^{(k)} = \hat{c}_{\text{AIR}}(\wt{F}^{(k)}, \wt{F}^{(k-1)}, \tilde{\pi}^{(k)}, \tilde{\pi}^{(k-1)}) \times c_{\text{IS}}^{(k)} \label{eqn:ISAIR-var-asym}
\end{align}
where $| \hat{c}_{\text{ISAIR}}^{(k)} - c_{\text{ISAIR}}^{(k)} | \to 0$ as $k \to \infty$. Just as for the ISMF estimator, we see a quasi-multiplicative effect, in that the asymptotic variance reduction $\hat{c}_{\text{ISAIR}}^{(k)}$ is the product of the variance reduction from IS alone, $c_{\text{IS}}^{(k)}$, with the functional form of the asymptotic variance reduction for AIR, $\hat{c}_{\text{AIR}}$, but now using $\wt{F}^{(k)}, \wt{F}^{(k-1)}, \tilde{\pi}^{(k)}, \tilde{\pi}^{(k-1)}$ instead of $F^{(k)}, F^{(k-1)}, \pi^{(k)}, \pi^{(k-1)}$.

\subsection{The ISMFAIR meta estimator: Importance sampling, adaptive information reuse, and multifidelity}

Lastly, we detail an estimator which simultaneously combines the MF, IS, and AIR method. To do this, we replace the MC estimators present in the AIR estimator (\ref{eqn:AIR-est}) with ISMF estimators. Equivalently, this can be viewed as introducing importance weights into the MFAIR scheme. For an equivalent budget of $p$ HFM samples, we define the importance sampled multifidelity adaptive information reuse estimator (ISMFAIR) as
\begin{align}
     \wh{F}_{\text{ISMFAIR},p}^{(k)} = \operatorname{MF}\lp \wt{F}^{(k)},\wt{G}^{(k)},\tilde{\pi}^{(k)}, \frac{p}{2} \rp + \gamma \lp \wh{F}_{\text{ISMFAIR},p}^{(k-1)} - \operatorname{MF} \lp \wt{F}^{(k-1)}W_{k},\wt{G}^{(k-1)},\tilde{\pi}^{(k)},\frac{p}{2}\rp \rp\,. \label{eqn:ISMFR-est}
\end{align}
Just as for the ISAIR estimator, the optimal $\gamma^{\ast}_{\text{ISMFAIR}}$ and variance of the ISMFAIR estimator are identical to that of the MFAIR estimator but with the relevent substitutions made. The optimal $\gamma^{\ast}_{\text{ISMFAIR}}$ is given by
\begin{align*}
    \gamma^{\ast}_{k,\text{ISMFAIR}} = \frac{\wt{\operatorname{COV}}_{k}}{\Var_{\tilde{\pi}^{(k-1)}}[\wh{F}_{\text{ISMFAIR},p}^{(k-1)}] + \Var_{\tilde{\pi}^{(k)}}[\wh{F}_{\text{ISMFAIR},p}^{(k-1)} - \wh{F}_{\text{MF},\wt{F}^{(k-1)}W_{k},\wt{G}^{(k-1)},\tilde{\pi}^{(k)},p/2}]} 
\end{align*}
where the term $\wt{\operatorname{COV}}_{k}$ is the covariance of $\wh{F}_{\text{MF},\wt{F}^{(k)},\wt{G}^{(k)},\tilde{\pi}^{(k)},p/2}$ and $\wh{F}_{\text{MF},\wt{F}^{(k-1)}W_{k},\wt{G}^{(k-1)},\tilde{\pi}^{(k)},p/2}$
under the distribution $\tilde{\pi}^{(k)}$. The formula for $\wt{\operatorname{COV}}_{k}$ is analogous to the formula for $\operatorname{COV}_{k}$, equation (\ref{eqn:MFIR-cov}), with all models replaced by their importance weighted counterparts and $\pi^{(k)}$ replaced by $\tilde{\pi}^{(k)}$.
The variance reduction of the ISMFAIR estimator $\wh{F}_{\text{ISMFAIR},p}^{(k)}$ using $\gamma^{\ast}_{k,\text{ISMFAIR}}$, compared to $\wh{F}_{\text{MC},p}^{(k)}$ is
\begin{align*}
    c_{\text{ISMFAIR}}^{(k)} = 
    \Var_{\pi^{(k)}}[\wh{F}_{\text{ISMFAIR},p}^{(k)}] / \Var_{\pi^{(k)}}[\wh{F}_{\text{MC},p}^{(k)}].
\end{align*}

Just as for the MFAIR estimator, the variance reduction $c_{\text{ISMFAIR}}^{(k)}$ depends on $c_{\text{ISMFAIR}}^{(k-1)}$ which subsequently depends on $c_{\text{ISMFAIR}}^{(k-2)}$ and so on. To avoid this recursive dependence, we derive we derive an asymptotic approximation for $c_{\text{ISMFAIR}}^{(k)}$, under similar assumptions as Proposition \ref{prop:MFIR-asym}, with the appropriate HFMs, surrogates, and distributions substituted in the assumptions.

This asymptotic approximation of $c_{\text{ISMFAIR}}^{(k)}$ is
\begin{align*}
    \hat{c}_{\text{ISMFAIR}}^{(k)} = \hat{c}_{\text{MFAIR}}(\wt{F}^{(k)}, \wt{F}^{(k-1)}, \wt{G}^{(k)}, \tilde{\pi}^{(k)}, \tilde{\pi}^{(k-1)}) \times c_{\text{IS}}^{(k)},
\end{align*}
where $|\hat{c}_{\text{ISMFAIR}}^{(k)} - c_{\text{ISMFAIR}}^{(k)}| \to 0$ as $k \to \infty$. Akin to our other meta estimators, we see that this asymptotic approximation of the variance reduction is again multiplicative in the sense that it is the product of the variance reduction from IS alone, $c_{\text{IS}}^{(k)}$, with the functional form of the asymptotic approximation of variance reduction for MFAIR, but using $\wt{F}^{(k)}, \wt{F}^{(k-1)}, \wt{G}^{(k)}, \tilde{\pi}^{(k)}, \tilde{\pi}^{(k-1)}$ instead of $F^{(k)}, F^{(k-1)}, G^{(k)}, \pi^{(k)}, \pi^{(k-1)}$.

\section{Energetic particles in stellarators} \label{section:physics}

We now apply the proposed meta estimator to efficiently estimate energetic particle confinement in stellarators. We first introduce the physical model of interest for the dynamics of energetic particles in fusion devices, and then present the application of the meta estimator for the confining quality of the magnetic field during optimization.

\subsection{Energetic particle dynamics and stellarator optimization}
We consider the dynamics of 3.5 MeV alpha particles born as a result of deuterium-tritium fusion in a three-dimensional stellarator magnetic field. Fusion reactions may be approximated as a probabilistic process with the following properties: the direction of the velocity of alpha particles at birth follows a uniform distribution, and if the deuterium and tritium nuclei are at the same uniform temperature throughout the domain, then the location of birth of alpha particles in that domain also follows a uniform distribution.

Studies of alpha particle confinement often rely on Monte Carlo estimators to properly capture the consequences of the probabilistic nature of the process \cite{henneberg2019properties,LandremanPaulPRL,Wechsung_2022,wechsung2022precise,giuliani_directcomputation,jorge2022single}. Deterministic measures of energetic particle confinement have been proposed for the design of magnetic field with good confinement properties \cite{Nemov2008,bader_2019,Bader_2021,Velasco_2021}, but their reliability and scope are limited \cite{Bader_2021,Velasco_2021}, due to the wide variety of particle orbits and of loss mechanisms \cite{Lotz_1992,Beidler_2001,Faustin_2016,Velasco_2021,White_2021,paul2022energetic}. Another strategy to achieve strong alpha particle confinement in stellarators without relying on expensive Monte Carlo estimation of confinement during the reactor optimization process is to obtain it as a natural by-product of another highly desirable property of the magnetic configuration targeted during optimization, called quasi-symmetry \cite{Helander_2014}, which also guarantees the confinement of the thermal deuterium-tritium fuel. It was indeed recently shown that magnetic fields with an unprecedented level of quasi-symmetry confine energetic alpha particles extremely well \cite{LandremanPaulPRL,wechsung2022precise,giuliani_directcomputation,landreman2022optimization,Wechsung_2022}. Based on these promising results, one could be tempted to conclude that energetic particle confinement codes need not be included in multi-physics stellarator design studies, and that good confinement should simply be verified numerically, via a single expensive Monte Carlo estimation, once an optimized configuration with good quasi-symmetry has been computed. There are two caveats to such a strategy. First, not all optimized stellarators have quasi-symmetric magnetic fields \cite{Subbotin_2006,Helander_2014,parra2015less,plunk_2019,jorge2022single}, and unlike quasi-symmetric fields, it has not yet been numerically shown that excellent energetic particle confinement naturally follows from the construction of these other types of optimized magnetic fields \cite{jorge2022single} . Second, the excellent confinement results found recently for quasi-symmetric configurations were obtained for designs that did not account for several engineering constraints and criteria, such as the geometry and location of plasma facing components \cite{greuner_2003}, and blanket design for thermal and neutral shielding as well as tritium breeding and heat exchange \cite{sorbom2015arc,bongiovi_blanket}. The level of quasi-symmetry of reactor designs accounting for these engineering constraints is likely to be lower than in the physics-driven designs recently published, as the optimization becomes more complex, and physics targets are no longer the only driving objectives. Since alpha particle confinement can degrade rapidly with increasing deviations from quasi-symmetry \cite{Bader_2021}, ensuring good confinement performance necessitates the inclusion of reliable measures of confinement in the optimization objectives, and thus the inclusion of Monte Carlo estimation, which is the most versatile method among the ones mentioned above.

In the Monte Carlo approach, one randomly selects initial conditions for the alpha particles corresponding to the probabilistic birth process described above, and then numerically integrates their trajectories to determine the fraction of particles in the sample that is eventually lost, as well as the average confinement time for that sample. In principle, these trajectories should account for the effect of collisions with the thermalized electrons and deuterium and tritium ions \cite{henneberg2019properties,Faustin_2016,Bader_2021,Lazerson_2021}. In practice however, ignoring the effect of collisions can still provide good accuracy for the energetic particle loss estimates, for times of flight up to a large fraction of the alpha particle slowing down time due to collisions \cite{Mynick_2006,Bader_2021,Lazerson_2021}. Since energetic particle losses occurring before their characteristic slowing down time are the most detrimental for a fusion nuclear power plant \cite{Mau_2008,Faustin_2016,bader_2019,Bader_2021,Velasco_2021}, from both power balance and material damage perspectives, and since collisionless orbits are easier and less computationally expensive to integrate than orbits including collisions, many energetic particle confinement studies for stellarator optimization are done based on collisionless orbits \cite{Helander_2014,Velasco_2021,LandremanPaulPRL,wechsung2022precise,giuliani_directcomputation,jorge2022single,Wechsung_2022}. This is also what we do in this article. We however stress that the methods we present here also apply to orbit integrators which are able to account for collisions. The analysis of the gains in efficiency provided by our methods when collisions are taken into account is left for future work.

\subsection{Models of collisionless dynamics of energetic particles}
The full collisionless dynamics of energetic particles born in stellarators is governed by the Lorentz force, according to Newton's second law of motion: $md\vb{v}/dt=q\vb{v}\times\vb{B}$, where $m$ is the particle mass, $q$ the particle charge, and $\vb{B}$ the magnetic field at the particle location. Computing the orbits given by these ordinary differential equations (ODEs) for the full extent of the slowing down time scale is computationally expensive, due to the multi-scale nature of the motion: to lowest order, the particles execute a fast quasi-helical motion centered on a magnetic field line, but the loss of confinement is due to the small departure of this motion from a perfect helix, which is called drift, and occurs on a much slower time scale \cite{Taylor_1964,Helander_2014,Velasco_2021}. To this day, in the absence of numerical ODE integrators capable of relying on the scale separation between the two types of motion to accelerate the computation of the particle orbits without a significant loss of accuracy, it remains intractable to include Monte Carlo simulations based on the full Newton's equations in stellarator optimization and design studies. To address this challenge, physicists have relied on a multiple time scale analysis to derive \textit{guiding center equations} \cite{Helander_2014}, which arise from averaging the equations of motion given by the Lorentz force over the fast helical motion \cite{Taylor_1964,helander2005collisional}, and which describe the motion of the average particle location during its helical motion, called the guiding center. In the limit in which the radius of the particle helical motion is negligible compared to the typical length scale of variation of the magnetic field, the guiding center equations provide an excellent approximation of the exact particle motion \cite{Helander_2014,Bader_2021}. This regime, which is observed for strong magnetic fields, is the regime of interest for magnetic fusion reactors. The accuracy of the guiding-center orbits in that regime has been verified numerically \cite{Lazerson_2021NF}. Since the guiding center equations are much less computationally expensive to integrate than the full Newton's equations, they are most commonly used for stellarator optimization \cite{ku2006new,bader_2019,Bader_2021,albert_2020,LandremanPaulPRL,wechsung2022precise,giuliani_directcomputation,paul2022energetic,Wechsung_2022}. We therefore also apply our variance reduction framework to the guiding center equations in this work, and not to the full Newton's equations. It has been recognized that the guiding center equations may have limited accuracy in a few situations of interest for reactor design \cite{Bader_2021,Liu_2021}. We note that all the variance reduction methods discussed in this paper can also be applied to full orbit dynamics. 

The guiding center equations can be represented as a four-dimensional system of ordinary differential equations (ODEs) corresponding to three spatial dimensions, and one dimension for the parallel velocity $v$. For the vacuum fields we will consider for our numerical tests in the next section, these dynamics are given by \cite{freidberg2008plasma}
\begin{align}
    \dot{\vb{x}} &= v \frac{\vb{B}^{(k)}}{B^{(k)}} + \frac{m}{q(B^{(k)})^{3}} \lp \frac{v_{\perp}^{2}}{2} + v^{2} \rp \vb{B}^{(k)} \times \nabla B^{(k)} \label{eqn:gc1} \\
    \dot{v} &= - \frac{\mu}{B^{(k)}} \vb{B}^{(k)} \cdot \nabla B^{(k)} \label{eqn:gc2}
\end{align}
where $m$ is the particle mass, $q$ is the particle charge, $\vb{x}$ is the position vector of the energetic particle, $\vb{B}^{(k)}$ is the magnetic field at the $k$th iteration determined by the outer-loop variables $\lambda^{(k)}$, $B^{(k)} = |\vb{B}^{(k)}(x,y,z)|$ is the field strength, $\mu$ is the magnetic moment \cite{Taylor_1964,Helander_2014}, and $v_{\perp}^{2} = 2\mu B^{(k)}$. Given an initial position  $\vb{x}(0) = \vb{x}_{0}$ and initial parallel velocity $v(0)=v_{0}$, energetic particles are traced by solving equations (\ref{eqn:gc1}-\ref{eqn:gc2}) until some final time $T_{\text{max}}$, such as the characteristic alpha particle slowing down time, or until they reach a closed flux surface we label as the plasma edge, and are considered lost. We stress once more that our methods are not limited to dynamics given by equations (\ref{eqn:gc1}-\ref{eqn:gc2}). These equations are highlighted here because they correspond to the situations we considered for our numerical examples.

As is common in many confinement studies \cite{henneberg2019properties,LandremanPaulPRL,wechsung2022precise,giuliani_directcomputation,Wechsung_2022}, we model particles as only being spawned on a single flux surface $S_{\text{spawn}}^{(k)}$ and consider particles lost when they reach a flux surface $S_{\text{exit}}^{(k)}$. The surfaces $S_{\text{spawn}}^{(k)}$ and $S_{\text{exit}}^{(k)}$ are chosen as the flux surfaces with \textit{fixed} flux label $s_{\text{spawn}}$ and $s_{\text{exit}}$ for the field $\vb{B}^{(k)}$, and as $\vb{B}^{(k)}$ changes, the spawn and exit surfaces themselves must change. Using a toroidal angle $\phi$ and poloidal angle $\theta$, the spawn surface $S_{\text{spawn}}^{(k)}$ can be described using with a function $\Gamma^{(k)}(\phi,\theta) = \vb{x}_{0}$, where $\Gamma^{(k)}:[0,1]^{2} \to S_{\text{spawn}}^{(k)}$ is a one-to-one correspondence. Assuming the deuterium and tritium nuclei are at the same uniform temperature on $S_{\text{spawn}}^{(k)}$, alpha particles must be born uniformly on $S_{\text{spawn}}^{(k)}$. As a result, we model the uncertainty in the particle birth distribution using the pullback distribution $\nu^{(k)} = (\Gamma^{(k)})^{-1}_{\#}\text{Unif}(S_{\text{spawn}}^{(k)})$ on the fixed angle space $[0,1]^{2}$. 

As discussed above, to model the uncertainty in initial birth velocity, we follow the convention that particles are born isotropically in 3D velocity space, so that the initial parallel velocity 
$v_{0}$ has distribution $q:=\text{Unif}(-V_{\text{max}}, V_{\text{max}})$ where $V_{\text{max}}$ is the speed of a particle born with kinetic energy 3.5 MeV. The input uncertainty we consider is then a random vector $\xi = (\phi,\theta,v_{0})$ on the domain $D = [0,1]^{2} \times [-V_{\text{max}}, V_{\text{max}}]$ whose distribution, at the $k$th outer-loop iteration, is given by the product distribution $\pi^{(k)} = \nu^{(k)} \times q$. For each outer-loop iteration $k$, we seek to estimate a metric of confinement $E_{\pi^{(k)}}[F^{(k)}]$ for the $k$th magnetic configuration, where $F^{(k)}(\xi)$ is given by the HFM describing the confinement of an energetic particle birthed with $\xi$. The confinement metric we consider in this work is the \textit{scaled mean modified lost time} \cite{Law2022}, where
\begin{align}
    F^{(k)}(\xi) = F^{(k)}(\phi,\theta,v_{0}) = \min(\inf\{t \, : \, \vb{x}(t) \in S_{\text{exit}}^{(k)}, \, \vb{x}(0) = \Gamma^{(k)}(\phi,\theta), v(0) = v_{0} \},T_{\text{max}})/T_{\text{max}}
\end{align}
where $\vb{x}(t) = (x(t),y(t),z(t))$ solves the dynamics (\ref{eqn:gc1}-\ref{eqn:gc2}) with magnetic field $\vb{B}^{(k)}$, initial position $\Gamma^{(k)}(\phi,\theta)$, and initial velocity $v_{0}$. As discussed in \cite{Law2022}, the modified loss time is a regularized version of the standard loss time, to address the mathematical difficulty that for confined particles, the time at which the particles are lost is infinite. To circumvent this singular behavior, we say that the loss time of confined particle is $T_{\text{max}}$. The outer loop application we consider is stellarator optimization, and the outer-loop variables $\lambda^{(k)}$ correspond to the Fourier coefficients of the stellarator coils and are the primary design optimization variables at hand. At each optimization iteration $k$, the $\lambda^{(k)}$ specify the stellarator coils which then controls the magnetic field $\vb{B}^{(k)}$ through the Biot-Savart law.

This field $\vb{B}^{(k)}$ then dictates the spawn surface $S_{\text{spawn}}^{(k)}$, the exit surface $S_{\text{exit}}^{(k)}$, as well as the particle dynamics (\ref{eqn:gc1}-\ref{eqn:gc2}). All three of these determine the HFM $F^{(k)}$. The spawn surface $S_{\text{spawn}}^{(k)}$ specifies the one-to-one correspondence $\Gamma^{(k)}$ which then determines $\nu^{(k)}$ and thus $\pi^{(k)}$. A diagram summarizing the dependencies is displayed in Figure \ref{fig:flowchart}.

\tikzstyle{basic} = [rectangle, rounded corners, minimum width=3.5cm, minimum height=1.5cm,text centered, draw=black,text width=3.5cm]
\tikzstyle{arrow} = [thick,->,>=stealth]

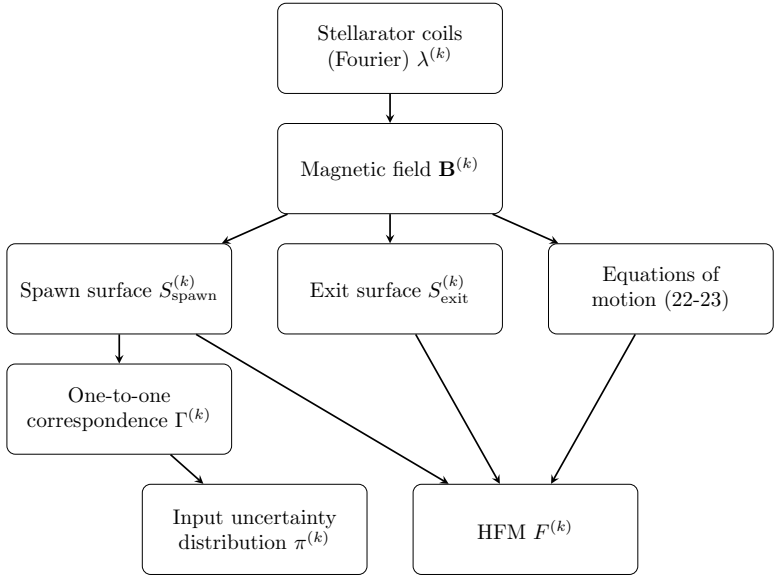
\begin{figure}[!t]
    \centering
    \scalebox{0.8}{
    \begin{tikzpicture}[node distance=2cm]
    \node (start) [basic] {Stellarator coils \\ (Fourier) $\lambda^{(k)}$};
    \node (B) [basic,below of=start] {Magnetic field $\vb{B}^{(k)}$};
    \node (exit) [basic,below of=B] {Exit surface $S_{\text{exit}}^{(k)}$};
    \node (dynamics) [basic,right of=exit, xshift=2.5cm] {Equations of \\motion (\ref{eqn:gc1}-\ref{eqn:gc2})};
    \node (spawn) [basic,left of=exit, xshift=-2.5cm] {Spawn surface $S_{\text{spawn}}^{(k)}$};
    \node (gamma) [basic,below of=spawn] {One-to-one \\ correspondence $\Gamma^{(k)}$};
    \node (pi) [basic,below of=gamma,xshift=2.25cm] {Input uncertainty distribution $\pi^{(k)}$};
    \node (F) [basic,below of=dynamics, yshift=-2cm,xshift=-2.25cm] {HFM $F^{(k)}$};
    
    \draw [arrow] (start) -- (B);
    \draw [arrow] (B) -- (exit);
    \draw [arrow] (B) -- (dynamics);
    \draw [arrow] (B) -- (spawn);
    \draw [arrow] (spawn) -- (gamma);
    \draw [arrow] (gamma) -- (pi);
    \draw [arrow] (dynamics) -- (F);
    \draw [arrow] (spawn) -- (F);
    \draw [arrow] (exit) -- (F);
    \end{tikzpicture}
    }
    \caption{Flow chart demonstrating how outer-loop variables $\lambda^{(k)}$ affect the input distribution $\pi^{(k)}$ and the HFM $F^{(k)}$ for the problem of energetic particle confinement.}
    \label{fig:flowchart}
\end{figure}

\subsection{Data-driven surrogate model for multifidelity estimation}

We now discuss our choice of surrogate model $G$ which we utilize in the four MF-based estimators (MF, ISMF, MFAIR, and ISMFAIR estimators). The primary reason we focus on constructing a data-driven surrogate models is that traditional sources of surrogate modeling are inappropriate for energetic particle motion \cite{Law2022}. For example, there is no hierarchy of simplified physics models to be leveraged \cite{Peherstorfer2018-MFreview}, since the guiding center model already results from gyro-averaging the true energetic particle dynamics determined by the Lorentz force. Moreover, the multi-scale nature of alpha particle dynamics means that surrogates arising from coarser time steps are also unreliable. Additionally, popular projection based model reduction techniques such as proper orthogonal decomposition are largely ineffective since the problem is transport based \cite{PeherstoAMS}.

For this purpose we utilize a data-driven surrogate model, namely an interpolant, which have served as surrogates in other works on multifidelity methods already \cite{Law2022}. We utilize trigonometric interpolation in the angle variables $(\phi,\theta)$ and piece-wise linear interpolation in the velocity variable $v_{0}$. We also leverage the domain knowledge that particles born with large parallel velocity $|v_{0}|$ are typically well confined, and thus we build our surrogate $G$ on a truncated subdomain of $[-V_{\text{max}},V_{\text{max}}]$. 

More specifically, we select a subinterval $I_\text{interpolate} \subset [-V_{\text{max}}, V_{\text{max}}]$. Within this subinterval, we construct $G$ to be an interpolant of $F^{(0)}$ using training-target pairs $\{(\xi_{i},F^{(0)}(\xi_{i})\}_{i=1}^{N_{\text{train}}}$ on the domain $[0,1]^{2} \times I_{\text{interpolate}}$. For $v_{0}$ outside $I_{\text{interpolate}}$, we set $G$ to be the constant value $T_{\text{max}}$. The choice of $I_{\text{interpolate}}$ is based on the discretion of the user and can be sourced from domain knowledge or pilot studies involving tracing a small number of particle trajectories. Further details on the interpolation used in our experiments is provided in Section \ref{section:numerics}.

\subsection{Data-driven biasing density for importance sampling}

While there are multiple mechanisms for particle loss in stellarators \cite{Lotz_1992,Beidler_2001,Faustin_2016,Velasco_2021,White_2021,paul2022energetic}, certain classes of particles are significantly more at risk of escaping confinement than others. Namely, passing particles which circulate rapidly around the stellarator are generally better confined than ``trapped'' particles which do not circulate around the entire device but instead bounce between two points of equal magnetic field strength and may slowly drift outwards \cite{Helander_2014}. These classes of orbits typically correspond to particles with larger $|v_{0}|$ (passing) and smaller $|v_{0}|$ (trapped). 

In order to leverage this domain knowledge, we construct a biasing density of the form $\tilde{\pi}^{(k)} = \nu^{(k)} \times \tilde{q}$ where $\tilde{q}$ is designed to capture this rough correspondence between orbits and $|v_{0}|$. We use a Gaussian mixture model (GMM) with two components to construct $\tilde{q}$ using training data from the initial configuration
\begin{align*}
    \tilde{q}(v_{0}) = w_{1} \scriptn(m_{1},\sigma_{1}^{2}) + w_{2} \scriptn (m_{2},\sigma_{2}^{2}), \quad w_{1} + w_{2} = 1\,.
\end{align*}

The variance of importance sampling estimators can be unbounded if the tails of the biasing density $\tilde{q}$ are not sufficiently heavy. vIn practice, it is often desirable for $\tilde{q}$ to have significantly heavier tails than $q$ to avoid numerical issues in unbalanced weights $q/\tilde{q}$. Thus, for safety, we multiply all GMM component variances by a safety factor once the GMM is trained.

Since our biasing density $\pi^{(k)}$ is the product distribution of $\nu^{(k)}$ for the spawn surface and $\tilde{q}$ for the parallel velocity, we see that our importance weights $\pi^{(k)} / \tilde{\pi}^{(k)} = q / \tilde{q}$ are independent of $k$. Since we are using the same surrogate model $G$ at every optimization iteration, then the importance weighted surrogate $\tilde{G} = G q / \tilde{q}$ is also independent of $k$. Further details of how we train our GMM for our numerical experiments are provided in Section \ref{section:numerics}.

\begin{remark}
Since our biasing density is designed to put mass on regions of $v_{0}$ which are more likely to be lost, this corresponds to regions where $F^{(k)}$ is smaller. Since we want $\pi^{(k)}$ to place mass on regions which contribute the most to $\E_{\pi}^{(k)}[F^{(k)}]$, for all IS-based estimators, namely the IS, ISMF, ISAIR, and ISMFAIR estimators, the HFM used is instead $1-F^{(k)}$. That is, we estimate $\E_{\pi^{(k)}}[1-F^{(k)}]$ directly and then use $\E_{\pi^{(k)}}[F^{(k)}] = 1 - \E_{\pi^{(k)}}[1-F^{(k)}]$. This does not change the variance reduction formulas, since using the MC estimator of $\E_{\pi^{(k)}}[F^{(k)}]$ and $\E_{\pi^{(k)}}[1-F^{(k)}]$ have the same variance, $\Var_{\pi^{(k)}}[1-F^{(k)}] = \Var_{\pi^{(k)}}[F^{(k)}]$.
\end{remark}

\begin{remark}
In this application, we construct our data-driven biasing density from the initial configuration, just as for our surrogate model. However, in principle one could adapt the biasing density based on the outer-loop iteration $k$. We note that there has been work done in adapting the biasing density during the outer-loop, such as reusing samples from previous iterations to construct an optimal biasing density for the current iteration \cite{Chaudhuri2020}. However, we leave the incorporation of adaptive biasing densities in meta estimators for future work.
\end{remark}

\begin{remark}
Since the surrogate $G$ only evaluates the interpolant within a subdomain  $I_\text{interpolate} \subset[-V_{\text{max}}, V_{\text{max}}]$, $G$ is significantly faster for particles which are spawned with $v_{0}$ outside $I_\text{interpolate}$. As a result, we note that $G$ is \textit{slower} to evaluate under the biasing distribution $\tilde{\pi}^{(k)}$, since more particles will be born in this subdomain under $\tilde{\pi}^{(k)}$ compared to $\pi^{(k)}$ which means more interpolant evaluations will be needed as compared to sampling from $\pi^{(k)}$. This leads to a different cost ratio $\tilde{w}$ in the ISMF and ISMFAIR estimators compared to the cost ratio $w$ in the MF and MFAIR estimators. \label{remark:speed}
\end{remark}

\section{Numerical results} \label{section:numerics}

We demonstrate the proposed meta variance reduction estimators on two stellarator configurations. The first configuration is inspired by the National Compact Stellarator Experiment (NCSX) \cite{Giuliani_nearaxis}. In this experiment, we only consider the magnetic field at a single optimization iteration, without taking the outer loop into account. As we will discuss in more detail below, this experiment serves to demonstrate our estimator in situations for which the magnetic field is not well optimized. In the second experiment, we demonstrate our meta estimators on an optimization trajectory of a quasi-axisymmetric  stellarator configuration, henceforth referred to as LPQA2022 \cite{LandremanPaulPRL}.

\subsection{Numerical setup}
We now provide details about the numerical setup.

\subsubsection{Properties of stellarator configurations}
The first configuration corresponds to the set of non-planar coils of NCSX. NCSX is a compact high performance stellarator which was designed in the 1990s and early
2000s to have a magnetic field which approximates quasi-axisymmetry \cite{reiman2001recent,zarnstorff2001physics}. The NCSX stellarator is composed of three unique modular coil shapes to which stellarator symmetry \cite{dewar1998stellarator} and three-fold toroidal symmetry are applied \cite{Giuliani_nearaxis}.
The NCSX design also relies on planar toroidal field coils and poloidal field coils, which are not included in our study. As a result, our set of non-planar coils generates a magnetic field with significant departures from quasi-symmetry \cite{Giuliani_nearaxis}, and relatively poor confinement properties. The point of this numerical example is to study the validity and robustness of our meta estimators without information reuse and in situations in which the magnetic field is not well optimized, which can happen in the first few iterations of an optimization study.

The second configuration that we consider is a coil system optimized to approximate the magnetic field with excellent quasi-symmetry and confinement properties designed by Landreman and Paul \cite{LandremanPaulPRL}. It was recently shown that excellent approximations of this remarkable field could be generated by a realistic set of electromagnets \cite{wechsung2022precise}. The authors of this work obtained this coil system by constructing an optimization problem\footnote{The details of the optimization problem can be found at https://github.com/fredglaw/meta-multifidelity.} relying on the approach of the FOCUS coil design tool \cite{Zhu_2017} and on several methods of the stellarator optimization code SIMSOPT \cite{Landreman_simsopt}. The purpose of this numerical example is two-fold. First, we are interested in the performance of our meta estimators for situations in which particles are very well confined, in contrast to our first numerical example. This example is therefore more relevant to the late stages of an stellarator optimization study. The second purpose of this numerical example is to investigate the capabilities of our meta estimators as it is applied at consecutive iterations of a design and optimization study.

\begin{table}
    \centering
    \begin{tabular}{|c|c|c|c|c|}
    \hline
    Configuration & $N_{\phi}$ & $N_{\theta}$ & $N_{v}$ & $I_{\text{interpolate}}$ \\
    \hline
    NCSX-like & 40 & 40 & 40 & $[-0.3 V_{\text{max}}, 0.3 V_{\text{max}}]$ \\
    LPQA2022 & 20 & 20 & 40 & $[-0.25 V_{\text{max}}, 0.45 V_{\text{max}}]$ \\
    \hline
    \end{tabular}
    \caption{Number of training points for interpolant $G$ as well as the subdomain of training for the parallel velocity.}
    \label{tab:surrogate-training}
    \vspace{2em}
    \centering
    \begin{tabular}{|c|c|c|c|c|c|c|}
    \hline
    Configuration & $m_{1}$ & $m_{2}$ & $\sigma_{1}^{2}$ & $\sigma_{2}^{2}$ & $w_{1}$ & $w_{2}$ \\
    \hline
    NCSX-like & $4.084 \times 10^{-2}$  & 
    $-3.632 \times 10^{-2}$ & $1.147 \times 10^{-2}$ & $8.048 \times 10^{-3}$ & $0.4994$ & $0.5006$\\
    LPQA2022 & $0.1406$ & $0.2024$ & $2.107 \times 10^{-3}$ & $3.376 \times 10^{-3}$ & $0.4846$ & $0.5154$\\
    \hline
    \end{tabular}
    \caption{Means, variances, and weights for trained GMM biasing densities using 5000 training-target pairs. The training data is scaled to lie in $[-1,1]$.}
    \label{tab:gmm-data}
\end{table}

\subsubsection{Setup of numerical solver}
In both coil systems we consider, the magnetic field is directly calculated from the coil geometry and coil currents, via the Biot-Savart law. For the coils, we rely on the common approximation that they are zero-thickness current-carrying filaments. Each filament is represented as a closed smooth curve in three dimensional space, which is described by a truncated Fourier series\footnote{The Fourier coefficients and the currents for the different coils for each example can be found at https://github.com/fredglaw/meta-multifidelity.} \cite{Zhu_2017,Giuliani_nearaxis,Wechsung_2022}.  Since the magnetic field is directly calculated from the Biot-Savart law, the existence of flux surfaces is not guaranteed, and not assumed. However, for the magnetic configurations we have considered, we were able to identify flux surfaces, and for fixed flux labels $s_{\text{spawn}}$ and $s_{\text{exit}}$, we numerically construct $S_{\text{spawn}}^{(k)}$ and $S_{\text{exit}}^{(k)}$ with the method described in \cite{giuliani_directcomputation}. For the NCSX-like coil set we use $s_{\text{spawn}} = 0.005$ and $s_{\text{exit}} = 0.3$, and for LPQA2022 we use $s_{\text{spawn}} = 0.1$ and $s_{\text{exit}} = 0.3$.

Particles are birthed in the angle space $[0,1]^{2}$ according to the pullback distribution $\nu^{(k)}$ discussed in Section \ref{section:physics}. Samples from $\nu^{(k)}$ are drawn by rejection sampling using a uniform proposal, where the target distribution is proportional to the Jacobian $\det (\partial \Gamma^{(k)} /\partial(\phi,\theta))$. The normalizing constant for this target density is computed to sufficiently high accuracy using trapezoidal quadrature for the doubly periodic $\Gamma^{(k)}$. Particle parallel velocity is sampled either uniformly according to $q$ or according to the GMM $\tilde{q}$.

\begin{figure}[!t]
    \centering
    \includegraphics[width=\textwidth]{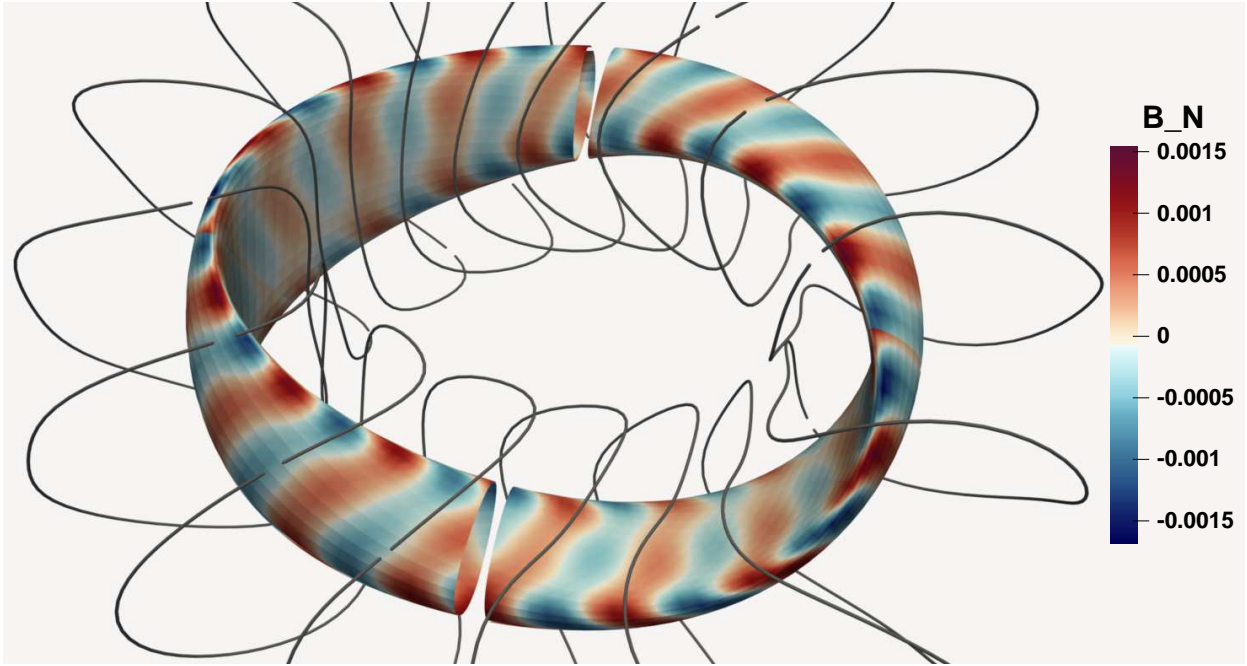}
    \caption{Final surface $S_{\text{exit}}^{(1000)}$ from the coil optimization with LPQA2022 as the target configuration. The $\vb{B}$ field is O(1) Tesla, so the normal component of the $\vb{B}$ field vanishes within 3 digit accuracy.}
    \label{fig:surface}
\end{figure}

Given initial conditions, particles are traced by solving the guiding-center equations (\ref{eqn:gc1}-\ref{eqn:gc2}) using an adaptive ODE integrator, namely Dormand-Prince, and followed until either time $T_{\text{max}} = 10^{-3}$ s or until they reach the exit surface $S_{\text{exit}}^{(k)}$ and are classified as lost. Classification of loss is done by tracking particles with a signed distance function which is positive in the volume contained by the exit surface $S_{\text{exit}}^{(k)}$, zero on the exit surface, and negative outside the exit surface.

We note that when integrating equations (\ref{eqn:gc1}-\ref{eqn:gc2}), we use a polynomial interpolant of the Biot-Savart magnetic field which is precomputed on a mesh in cylindrical coordinates. The interpolated magnetic field will not be exactly divergence free, however we ensure that the interpolation error is sufficiently small compared to the error of the numerical integration.

For all our numerical tests, we use protons with 9 keV of kinetic energy as proxies for alpha particles with 3.5 MeV in a reactor scale device with the dimensions of the ARIES-CS stellarator power plant design \cite{najmabadiARIES,lazersonproton,Wechsung_2022}. Specifically, if we call $\rho_{\star}$ the ratio of the small radius of the helical motion of particles around field lines to the characteristic radius of the plasma cross section, then with a kinetic energy of 9 keV, the proton trajectories we compute have the same $\rho_{\star}$ in the magnetic configurations we consider in this manuscript as 3.5 MeV alpha particles in the ARIES-CS device. Since $\rho_{\star}$ is the key non-dimensional parameter of interest for particle transport, we thus expect our results to be relevant for fusion power plant devices. As a consequence, physical parameters in our numerical experiments, such as maximum velocity, mass, charge, and field strength, are scaled for protons.

Our numerical examples were implemented in the SIMSOPT package \cite{Landreman_simsopt}. Further details on the coil parameterizations and currents, the Biot-Savart evaluations, the particle trajectory integrator, and the classifier for the numerical examples can be found at https://github.com/hiddenSymmetries/simsopt.

\subsubsection{Data-driven surrogate models}
For our data-driven surrogate, we utilize trigonometric interpolation in the angle variables $(\phi,\theta)$ and piece-wise linear interpolation in the velocity variable $v_{0}$ on the domain $[0,1] \times [0,1] \times I_{\text{interpolate}}$. This is done using $N_{\phi}$, $N_{\theta}$, $N_{v}$ equispaced points for $\phi, \theta, v_{0}$ respectively. The number of grid points and the parallel velocity subdomain used to train the surrogate for each test configuration is provided in Table \ref{tab:surrogate-training}. Each subdomain was chosen heuristically by examining a pilot study of 1000 particle trajectories. While more systematic choices of $I_{\text{interpolate}}$ are certainly possible, we utilize this ad-hoc choice since the focus in this work is on the efficacy of meta estimators provided $G$ and not the construction of $G$ itself. 

\begin{figure}[t]
    \centering
    \includegraphics[width=0.49\textwidth]{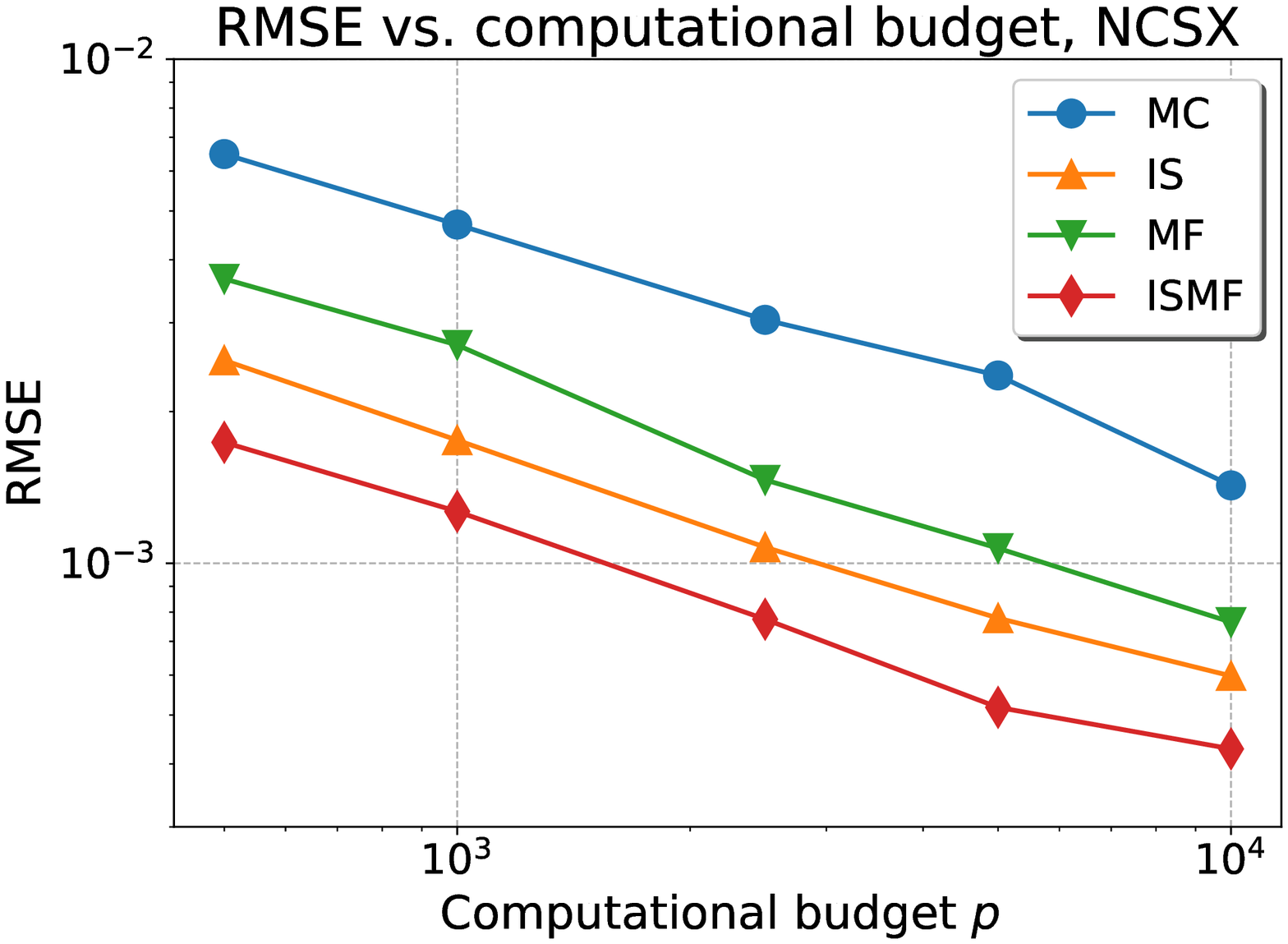}
    \includegraphics[width=0.49\textwidth]{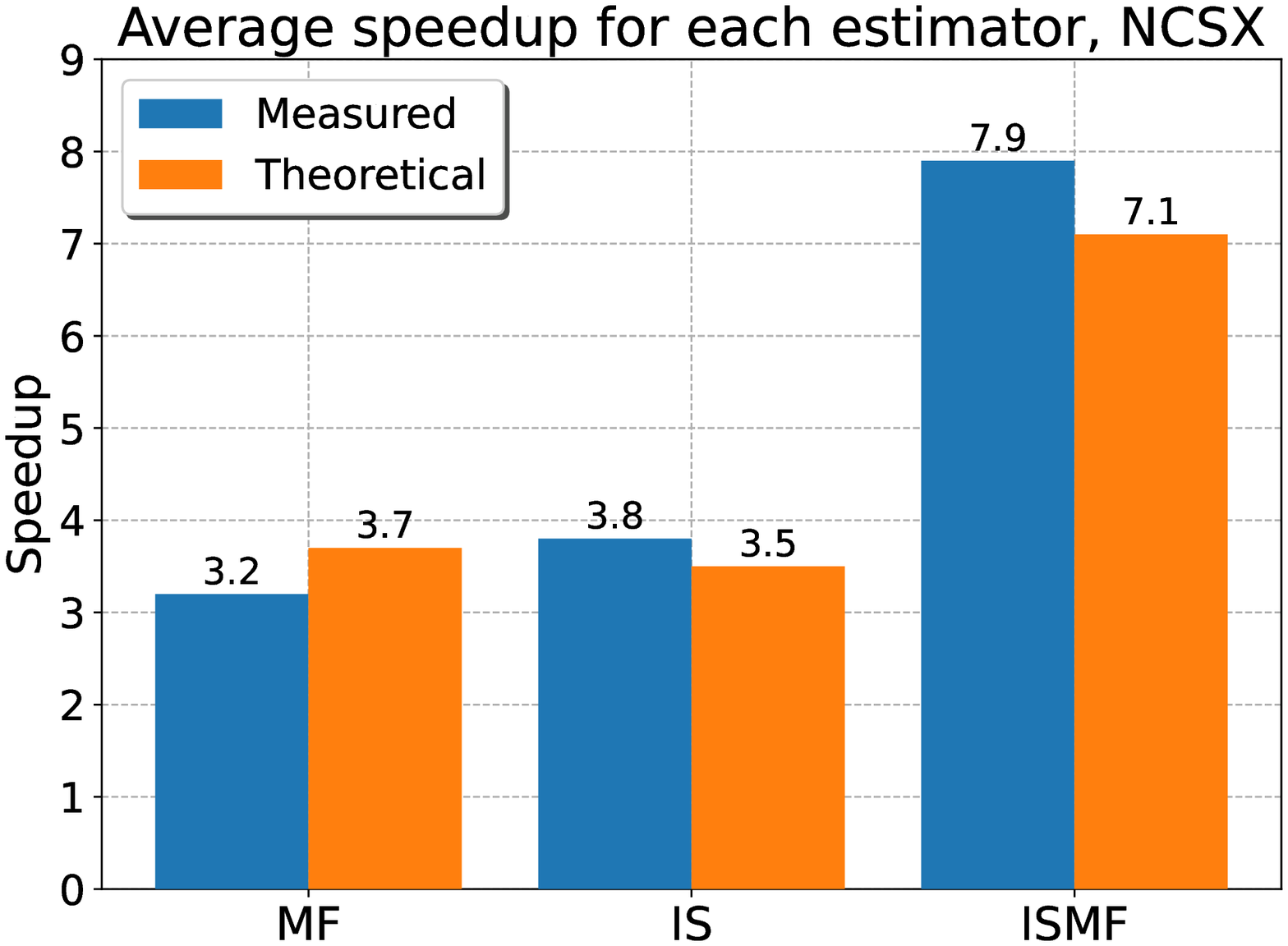}
    \caption{\textbf{Left:} RMSE versus computational budget $p$ for the NCSX-like configuration. While both IS and MF provide variance reduction on their own, the combined MFIS estimator improves on both. \textbf{Right:} Measured versus theoretical speedup for the MF, IS and ISMF estimators. Theoretical speedup for IS estimator is estimated using 5000 samples.}
    \label{fig:RMSE-vs-budget-NCSX}
\end{figure}

For our GMM $\tilde{q}$, we generate training data by drawing samples of $v_{0}$ from a centered Gaussian with most of its mass concentrated near 0. We choose this initial Gaussian to be intentionally heavy-tailed as we simply want to generate training data that is reflective of the domain knowledge we are leveraging.
We note that this is similar to the subinterval $I_{\text{interpolate}}$ utilized by our surrogate model, as both are drawing from the same source of domain knowledge. The distinction however is that for the GMM we simply use this as a guide to generate the training data itself. In this application we choose the initial Gaussian to place 95\% of its mass on the interval $[-0.3V_{\text{max}},0.3 V_{\text{max}}]$. We train our Gaussian using the expectation maximization algorithm in Scikit-learn \cite{scikit-learn} using 5000 training-target pairs, where the training data is drawn i.i.d. from the product distribution of $\nu^{(k)}$ and this initial Gaussian, and the targets are the classification of whether a particle is lost or not. The training pairs are scaled by $V_{\text{max}}$ to lie in $[-1,1]$. To avoid numerical blowup of our trained $\tilde{q}$, we multiple the standard deviation of each component by a safety factor of 2.25. This training procedure is the same for the NCSX and LPQA2022 cases, using training data from those cases respectively.

\begin{table}[!b]
    \centering
    \begin{tabular}{|c|c|c|}
    \hline
    Configuration & $w$ & $\tilde{w}$\\
    \hline
    NCSX-like & 355 & 54 \\
    LPQA2022 & 242 & 35 \\
    \hline
    \end{tabular}
    \caption{Approximate cost ratios $w$ and $\tilde{w}$ for the NCSX-like and LPQA2022 configurations. These values are estimates using 5000 samples for the initial configuration.}
    \label{tab:cost-ratios}
\end{table}

\begin{table}[!b]
    \centering
    \begin{tabular}{|c|c|c|c|c|c|}
    \hline
     \rule{0pt}{12pt} $k$  & $\Var_{\pi^{(k)}}(F^{(k)})$ & $\Var_{\pi^{(k)}}(F^{(k-1)}W_{k})$ & $\Var_{\tilde{\pi}^{(k)}}(\wt{F}^{(k)})$ & $\Var_{\tilde{\pi}^{(k)}}(\wt{F}^{(k-1)}W_{k})$ & $c_{\text{IS}}^{(k)}$\\
     \hline
     600 & 2.718e-02 & 2.705e-02 & 2.772e-03 & 2.774e-03 & 0.1020 \\
     700 & 2.723e-02 & 2.690e-02 & 2.755e-03 & 2.766e-03 & 0.1012 \\
     800 & 2.727e-02 & 2.722e-02 & 2.788e-03 & 2.766e-03 & 0.1022 \\
     900 & 2.673e-02 & 2.719e-02 & 2.763e-03 & 2.757e-03 & 0.1034 \\
     1000 & 2.722e-02 &  2.676e-02 & 2.757e-03 & 2.788e-03 & 0.1013 \\
     \hline
    \end{tabular}
    \caption{Measured variances and estimates of variance reduction using importance sampling for different $k$ in the LPQA2022 case. Each variance is the average of 150 replicates of sample variances, where each sample variance uses $p/2=2500$ samples. Each $c_{\text{IS}}^{(k)}$ is estimated as $\Var_{\tilde{\pi}^{(k)}}(\wt{F}^{(k)})/\Var_{\pi^{(k)}}(F^{(k)})$ using the measured variances in the table.}
    \label{tab:variances}
\end{table}

\subsection{Configuration 1: NCSX-like configuration}

We first compare the ISMF estimator against the MC, MF, and IS estimators on the magnetic configuration arising from the NCSX-like coil set. In this case, there is no outer-loop application, and thus we shall omit the notational dependence on $k$ for this case. As there is not outer-loop application, the primary purpose of this numerical test is to demonstrate the quasi-multiplicative variance reduction of our ISMF estimator (\ref{eqn:ISMF-var}). Using 1000 samples, we report a correlation of $\rho_{\pi}(F,G) \approx 0.8884$ and $\rho_{\tilde{\pi}}(\wt{F},\wt{G}) \approx 0.8171$. The cost ratios $w$ and $\tilde{w}$ are reported in Table \ref{tab:cost-ratios}, where we see that the cost ratio is 7 times smaller when sampling from $\tilde{\pi}$ compared to sampling from $\pi$. This is because our trained GMM $\tilde{q}$ has the bulk of its probability mass concentrated in $I_{\text{interpolate}}$, as can be seen in Table \ref{tab:gmm-data}. As a result, when sampling the parallel velocity from $\tilde{q}$, more samples land in $I_{\text{interpolate}}$, leading to interpolant evaluation which is more costly than simply assigning $T_{\text{max}}$ as the output.

We note that when switching from using $F,G$ in the MF estimator to $\wt{F},\wt{G}$ in the ISMF estimator, both the correlation and cost ratios \textit{decrease}, which suggests that $c_{\text{MF}}(\wt{F},\wt{G},\tilde{\pi}) > c_{\text{MF}}(F,G,\pi)$. However, provided that the variance reduction $c_{\text{IS}}$ from the IS estimator is sufficiently small, we may expect the ISMF estimator to outperform the MF and IS estimators alone. 

Using computational budgets $p=500,1000,2500,5000,10000$ we generated 250 replicates of the MC, IS, MF, and ISMF estimators. In Figure \ref{fig:RMSE-vs-budget-NCSX} we plot the RMSE of each estimator as a function of the budget $p$ for the scaled mean modified loss time $E_{\pi}[F]$. We see that while both the MF and IS estimators provide variance reduction compared to the MC estimator, the combined ISMF estimator outperforms the constituent MF and IS estimators. Based on the empirically measured $c_{\text{IS}}$, correlations $\rho_{\pi}(F,G)$ and $\rho_{\tilde{\pi}}(\wt{F},\wt{G})$, and cost ratios in Table \ref{tab:cost-ratios}, our numerical results agree with theoretical estimates to leading order. Thus we indeed observe quasi-multiplicative speedup for our ISMF estimator.

\subsection{Configuration 2: LPQA2022 configuration}

We now test all our meta estimators on a coil optimization trajectory whose target configuration is the LPQA2022 configuration. Since this optimization trajectory starts with circular coils, we focus on estimating particle confinement in the second half of the trajectory. That is, we train our surrogate model $G$ and GMM $\tilde{q}$ starting at the $k=500$th optimization trajectory and test our meta estimators at the $k=600,700,800,900,1000$ optimization iterations.

In Table \ref{tab:cost-ratios} we see a similar trend in cost ratios as in the NCSX-like case when introducing the GMM $\tilde{q}$ for sampling the parallel velocity. The reasoning is the same as for the NCSX-like case: the trained $\tilde{q}$ places much of its mass on $I_{\text{interpolate}}$, which makes the surrogate model more expensive on average. We note that $I_{\text{interpolate}}$ is not centered around zero for the LPQA2022 case, as can be seen in Table \ref{tab:surrogate-training}. We empirically discovered that we should select $I_{\text{interpolate}}$ in this way during our pilot study. We also observe that although the data used to train our GMM was drawn from a centered Gaussian, the trained $\tilde{q}$ is both shifted and skewed to the right, see Table \ref{tab:gmm-data}. Thus the overlap between $I_{\text{interpolate}}$ for the surrogate and where $\tilde{q}$ places most of its mass was not entirely chosen by us. Rather, we selected an $I_{\text{interpolate}}$ and the expectation-maximization trained $\tilde{q}$ to place mass there as well.

\begin{figure}[!t]
    \centering
    \includegraphics[width=0.49\textwidth]{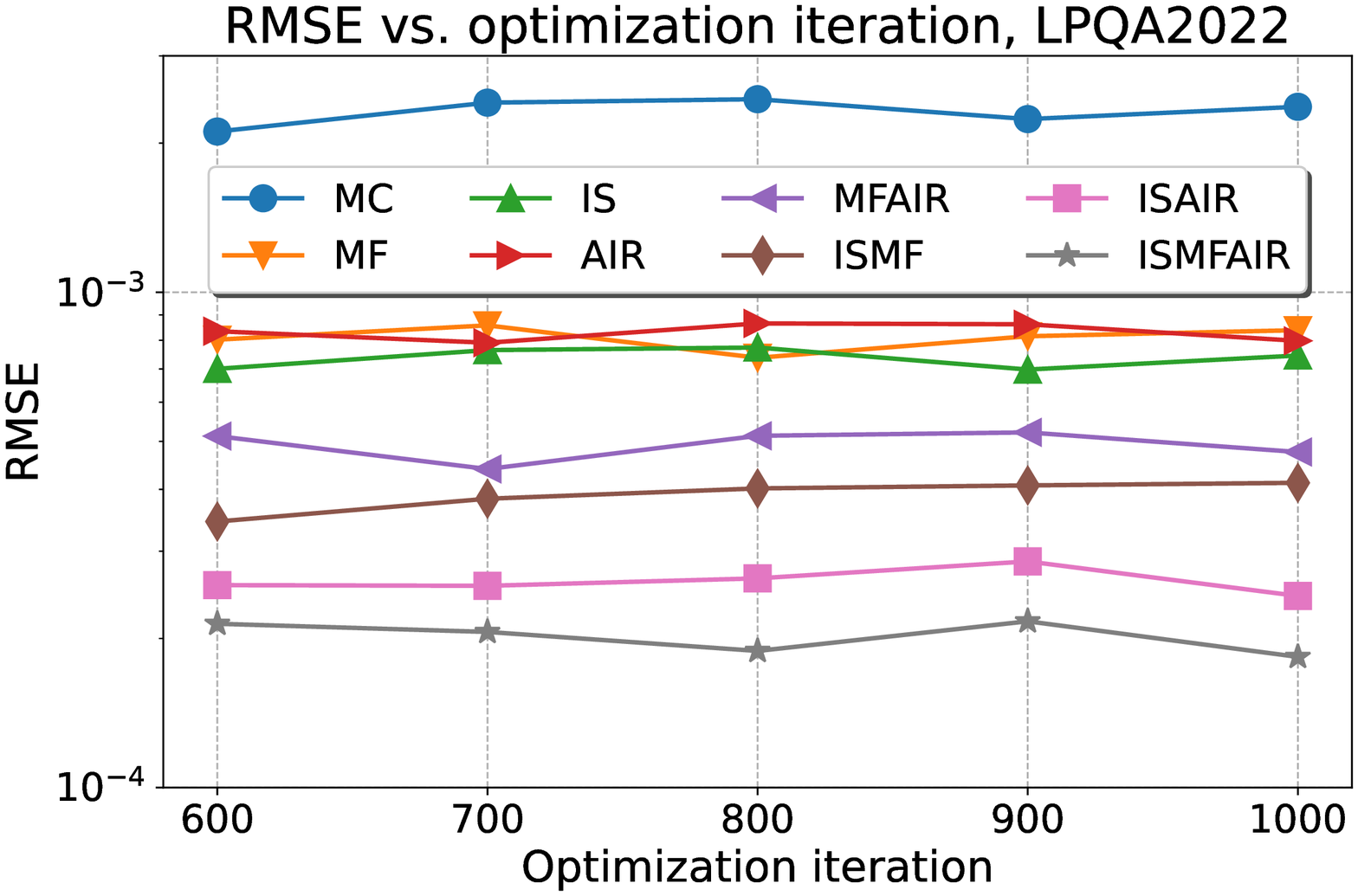}
    \includegraphics[width=0.49\textwidth]{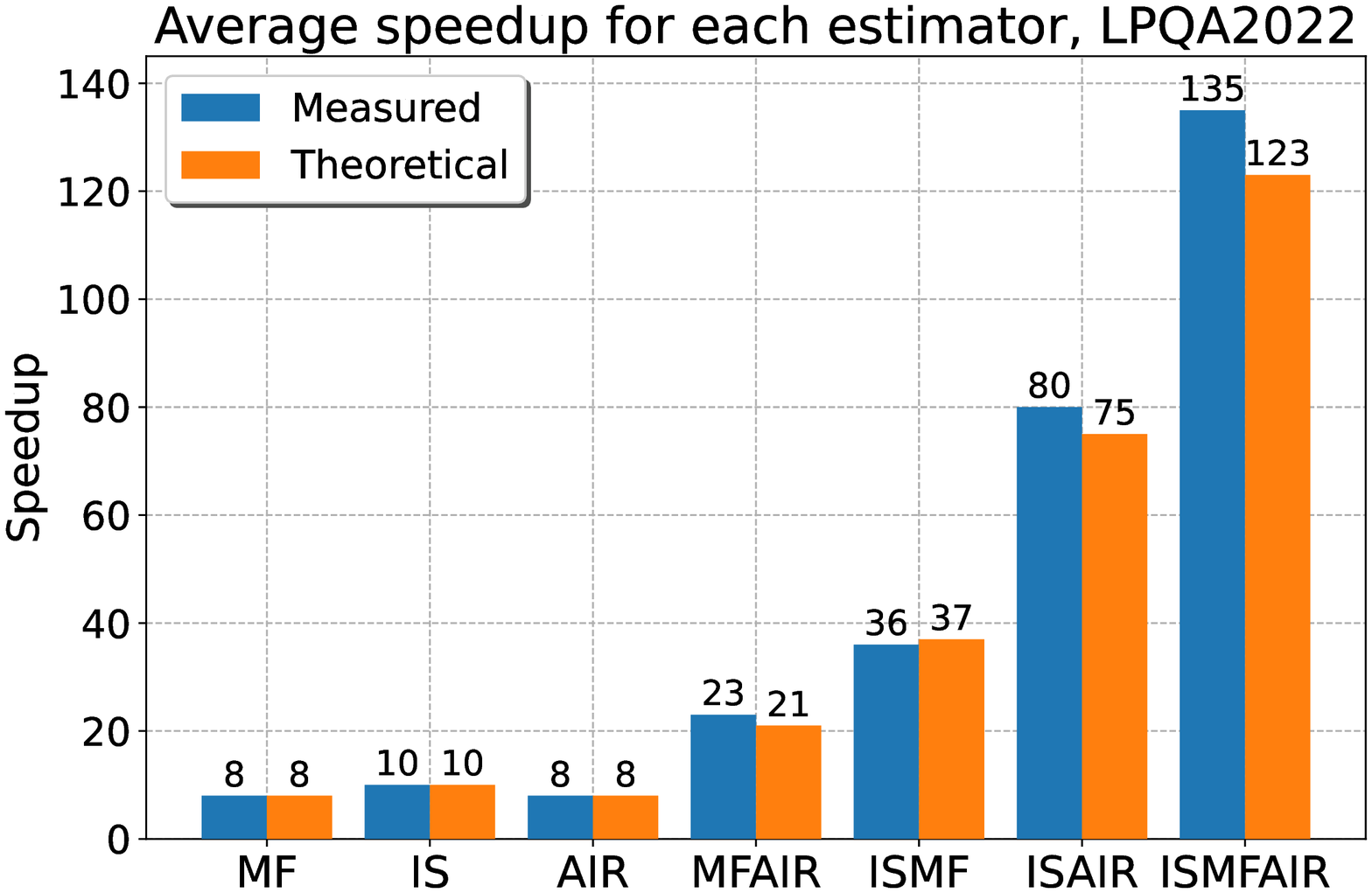}
    \caption{\textbf{Left:} Measured RMSE of each estimator at $k=600,700,800,900,1000$. Each variance is estimated using 150 estimator replicates, with each estimator using a budget of $p=5000$ HFM evaluations. \textbf{Right:} Measured and theoretical speedups for each method compared to the MC estimator with equivalent cost, averaged over their values at each $k=600,700,800,900,1000$. Each speedup is measured using 150 estimator replicates, where each estimator used a budget of $p=5000$ HFM evaluations. For each IR-based method, the theoretical speedup presented is based on the approximation of asymptotic variance reduction. Theoretical estimates at each $k$ are computed using $c_{\text{IS}}^{(k)}$ from Table \ref{tab:variances} and measured correlations in Table \ref{tab:correlations}. Bar plot labels are rounded to the nearest integer.}
    \label{fig:LPQA2022}
\end{figure}

Using a budget of $p=5000$, we generated 150 replicates of MC, MF, IS, AIR, MFAIR, ISMF, ISAIR, and ISMFAIR estimators. In Figure \ref{fig:LPQA2022} we plot the RMSE for each estimator for the scaled mean modified loss time $E_{\pi^{(k)}}[F^{(k)}]$ as a function of the optimization iteration $k$. Recall that although our surrogate and GMM were trained only using data at the $k=500$ iteration, the variance reduction for all estimators is practically constant for $k=600,700,800,900,1000$. This suggests that for late-stage optimization applications, data-driven surrogates and biasing densities may be highly effective even when not modified and adapted at each iteration. Moreover, we see that each meta estimator, i.e. MFAIR, ISMF, ISAIR, and ISMFAIR, outperforms its constituent estimators.

In Figure \ref{fig:LPQA2022} we plot the speedup of each constituent and meta estimator compared to the MC estimator, averaged over the values at $k=600,700,800,900,1000$. We also plot the theoretical speedup for the MF, IS, ISMF estimators as well as the asympotic speedup for the AIR, MFAIR, ISAIR, and ISMFAIR estimators. The theoretical and asymptotic values are estimated using the variances and correlations in Tables \ref{tab:variances} and \ref{tab:correlations} respectively, and then averaged over $k=600,700,800,900,1000$. We see excellent agreement between the measured speedups and the theoretical and asymptotic speedups for all of our estimators. For the ISMFAIR estimator which leverages all three of our constituent estimators, we measure over two orders of magnitude speedup compared to regular Monte Carlo.

Recall that in deriving our asymptotic variance reduction formulas, we relied on certain assumptions about the convergence of variances and correlations. In deriving the asymptotic variance reduction for the AIR estimator, we assumed that $\rho_{\pi^{(k)}}(F^{(k)},F^{(k-1)}W_{k})$ converged and that $\Var_{\pi^{(k-1)}}(F^{(k-1)}) / \Var_{\pi^{(k)}}(F^{(k-1)}W_{k}) \to 1$ as $k \to \infty$. Examining Tables \ref{tab:variances} and \ref{tab:correlations}, we approximately verify that this is indeed the case. Note that while we do not measure $\Var_{\pi^{(k-1)}}(F^{(k-1)})$, we observe that $\Var_{\pi^{(k)}}(F^{(k)})$ remains primarily constant in $k$, and thus deduce that $\Var_{\pi^{(k)}}(F^{(k)}) \approx \Var_{\pi^{(k-1)}}(F^{(k-1)})$ for large $k$. Similar assumptions hold for the asymptotic variance reduction of the ISAIR estimator with $\rho_{\tilde{\pi}^{(k)}}(\wt{F}^{(k)},\wt{F}^{(k-1)}W_{k})$ converging and with $\Var_{\tilde{\pi}^{(k-1)}}(\wt{F}^{(k-1)}) / \Var_{\tilde{\pi}^{(k)}}(\wt{F}^{(k-1)}W_{k}) \to 1$ as $k\to \infty$. In deriving the asymptotic variance reduction for MFAIR, we assumed that $\rho_{\pi^{(k)}}(F^{(k)},G)$ and $\rho_{\pi^{(k)}}(F^{(k-1)}W_{k},G)$ both converged to the same value as $k \to \infty$. Moreover, we assumed that $\rho_{\pi^{(k)}}(F^{(k)},F^{(k-1)}W_{k})$ converged as $k \to \infty$ and that $\Var_{\pi^{(k-1)}}(F^{(k-1)}) / \Var_{\pi^{(k)}}(F^{(k-1)}W_{k}) \to 1$ as $k \to \infty$. Examining the measured values in Tables \ref{tab:variances} and \ref{tab:correlations}, we can again approximately verify these assumptions. Likewise for the ISMFAIR estimator, we observe that $\rho_{\tilde{\pi}^{(k)}}(\wt{F}^{(k)},G)$ and $\rho_{\tilde{\pi}^{(k)}}(\wt{F}^{(k-1)}W_{k},G)$ roughly converge to the same value, $\rho_{\tilde{\pi}^{(k)}}(\wt{F}^{(k)},\wt{F}^{(k-1)}W_{k})$ seems to converge, and $\Var_{\pi^{(k-1)}}(F^{(k-1)}) / \Var_{\pi^{(k)}}(F^{(k-1)}W_{k}) \to 1$ as $k \to \infty$ is also approximately verified.

\begin{table}[!t]
    \centering
    \begin{tabular}{|c|c|c|c|}
    \hline
     \rule{0pt}{10pt} $k$ & $\rho_{\pi^{(k)}}(F^{(k)},G)$ & $\rho_{\pi^{(k)}}(F^{(k-1)}W_{k},G)$ & $\rho_{\pi^{(k)}}(F^{(k)},F^{(k-1)}W_{k})$\\
     \hline
     600 & 0.9580 & 0.9573 & 0.9980 \\
     700 & 0.9563 & 0.9561 & 0.9974 \\
     800 & 0.9571 & 0.9566 & 0.9981 \\
     900 & 0.9559 & 0.9569 & 0.9977 \\
     1000 & 0.9565 & 0.9560  & 0.9990 \\
     \hline
     \rule{0pt}{12pt}& $\rho_{\tilde{\pi}^{(k)}}(\wt{F}^{(k)},\wt{G})$ & $\rho_{\tilde{\pi}^{(k)}}(\wt{F}^{(k-1)}W_{k},\wt{G})$ & $\rho_{\tilde{\pi}^{(k)}}(\wt{F}^{(k)}, \wt{F}^{(k-1)} W_{k})$  \\
     \hline 
     600 & 0.9358 & 0.9384  & 0.9978 \\
     700 & 0.9343 & 0.9350 & 0.9971 \\
     800 & 0.9359 & 0.9351 & 0.9977 \\
     900 & 0.9360 & 0.9347 & 0.9974 \\
     1000 & 0.9346 & 0.9356 & 0.9988 \\
     \hline
    \end{tabular}
    \caption{Measured correlations between different models under different input distributions in the LPQA2022 case. Each correlation is estimated by averaging over 150 replicates of sample correlation. Sample correlations between high-fidelity were estimated using $p/2=2500$ samples, whereas sample correlations between high-fidelity and surrogate models were estimated using $n$ samples, where $n$ is the number of HFM evaluations dictated by the MF estimator with relevant models.}
    \label{tab:correlations}
\end{table}

\section{Summary and Discussion} \label{section:conclusion}

We have introduced meta multifidelity estimators which simultaneously leverage control variates, importance sampling, and information reuse for variance reduction. Our meta estimators provide quasi-multiplicative speedup, and for the AIR, MFAIR, ISAIR, and ISMFAIR estimators we have derived asymptotic approximations for variance reduction which can be computed without information from prior outer-loop iterations. We tested the performance of our meta estimators for estimating energetic particle confinement in stellarators during a stellarator optimization outer loop. Our numerical experiments demonstrated that our meta estimators outperform their constituent estimators, providing up to two orders of magnitude speedup compared to standard Monte Carlo estimation at equivalent computational cost.

Our new meta estimators are designed specifically for outer-loop applications such as stellarator optimization, and thus address a key issue raised in earlier work, namely the high cost of building the surrogate model initially. Since the surrogate model is built from the initial configuration, these meta estimators are particularly powerful for late-stage optimization, or in scenarios in which the initial configuration already has strong confinement but needs to be optimized for other desirable properties. In the latter case, the ability to directly estimate energetic particle confinement during optimization is critical for constrained optimization.

\section*{Acknowledgements}
The authors would like to thank the SIMSOPT development team, as well as David Pfefferl\'e for his insight on importance sampling for energetic particle confinement. Frederick Law was supported by the Department of Defense National Defense Science and Engineering Graduate Fellowship (DoD-NDSEG) and supported in part by the Research Training Group in Modeling and Simulation funded by the National Science Foundation via Grant RTG/DMS - 1646339. Antoine Cerfon was supported by the United States National Science Foundation under Grant No. PHY-1820852 and by the United States Department of Energy, Office of Fusion Energy Sciences, under Grant No. DE-FG02-86ER53223. Benjamin Peherstorfer was supported by the Air Force Office of Scientific Research (AFOSR)
award FA9550-21-1-0222 (Dr.~Fariba Fahroo) and the US Department of Energy, Office of Advanced Scientific Computing Research, Applied Mathematics Program (Program Manager Dr.~Steven Lee), DOE Award DESC0019334. This work was supported in part through the NYU IT High Performance Computing resources, services, and staff expertise.

\bibliographystyle{abbrv}
\bibliography{cas-refs}

\end{document}